\documentclass[fleqn,usenatbib]{mnras}
\usepackage[T1]{fontenc}
\DeclareRobustCommand{\VAN}[3]{#2}
\let\VANthebibliography\thebibliography
\def\thebibliography{\DeclareRobustCommand{\VAN}[3]{##3}\VANthebibliography}

\usepackage{graphicx}
\usepackage{amsmath}
\usepackage{amssymb}
\usepackage{siunitx}
\usepackage{multirow}
\usepackage{supertabular}

\usepackage{caption}

\DeclareSIUnit\Jy{Jy}
\DeclareSIUnit\pc{pc}
\DeclareSIUnit\yr{yr}
\DeclareSIUnit\erg{erg}

\title[CHIME FRB unsupervised machine learning]{Machine learning classification of CHIME fast radio bursts: II. Unsupervised Methods}
\author[J-M Zhu-Ge et. al]{Jia-Ming Zhu-Ge$^{1}$\thanks{zhugejiaming@mail.ustc.edu.cn}, Jia-Wei Luo$^{2,3}$ and Bing Zhang$^{2,3}$
\\
$^{1}$Department of Astronomy, University of Science and Technology of China, Hefei 230026, China\\
$^{2}$Nevada Center for Astrophysics, University of Nevada, Las Vegas, NV 89154, USA\\
$^{3}$Department of Physics and Astronomy, University of Nevada, Las Vegas, NV 89154, USA\\
}
\date{Accepted XXX. Received YYY; in original form ZZZ}

\pubyear{2022}

\begin{document}
\label{firstpage}
\pagerange{\pageref{firstpage}--\pageref{lastpage}}
\maketitle

\begin{abstract}
Fast radio bursts (FRBs) are one of the most mysterious astronomical transients. Observationally, they can be classified into repeaters and apparently non-repeaters. However, due to the lack of continuous observations, some apparently repeaters may have been incorrectly recognized as non-repeaters. In a series of two papers, we intend to solve such problem with machine learning. In this second paper of the series, we focus on an array of unsupervised machine learning methods. We apply multiple unsupervised machine learning algorithms to the first CHIME/FRB catalog to learn their features and classify FRBs into different clusters without any premise about the FRBs being repeaters or non-repeaters. These clusters reveal the differences between repeaters and non-repeaters. Then, by comparing with the identities of the FRBs in the observed classes, we evaluate the performance of various algorithms and analyze the physical meaning behind the results. Finally, we recommend a list of most credible repeater candidates as targets for future observing campaigns to search for repeated bursts in combination of the results presented in Paper I using supervised machine learning methods.
\end{abstract}


\section{Introduction}
\label{sec:introduction}

Fast Radio bursts (FRBs) are transient radio pulses that were first discovered by the Parkes Telescope in 2007 \citep[][]{2007Lorimer}. Most FRBs are in cosmological distances and last in millisecond time scales \citep[for reviews, see e.g.][]{2018Katz_FRB, Popov_2018, 2019Cordes_FRB, Petroff2019, PLATTS2019, 2020Nature_Zhang, Xiao2021, 2022Xiao, 2022Petroff_FRB}. Their observed properties, such as dispersion measure (DM)
and rotation measure (RM) give us insight into the physical properties (density, magnetic field strength) of the medium they pass through. The derived physical measures, such as energy or brightness temperature, aid to our understanding of the engine and radiation mechanism of the FRB sources. Although many theories have been proposed to explain these properties \citep[][and references therein]{PLATTS2019,2020Nature_Zhang, Xiao2021}, few of them have been comprehensively verified or successfully interpreted data in a universal way. 

FRBs are categorized into two types based on their repetition properties: repeaters and apparent non-repeaters. As their names suggest, repeaters are FRB sources with multiple bursts detected, while apparent non-repeaters have only been detected once. Since the striking discovery of the first repeater FRB 20121102A in 2016 \citep[][]{2016Spitler121102, Scholz_2016_121102}, up to now, about 2 dozens of repeaters have been reported. In the first CHIME/FRB catalog, 62 repeater bursts from 18 repeaters and 474 apparently non-repeating FRB bursts are listed \citep[][]{CHIME1}. Whether repeaters and apparent non-repeaters are distinct categories has been extensively discussed in the literature \citep{palaniswamy2018AreThereMultiple,caleb2019AreAllFast,ai2021TrueFractionsRepeating}. Some studies \citep{andersen2019CHIMEFRBDiscovery,fonseca2020NineNewRepeating,li2021BimodalBurstEnergy,aggarwal2021ObservationalEffectsBanded,CHIME1,pleunis2021FastRadioBurst,li2021LongShortFast,xiao2022NewInsightsCriterion,zhang2022StatisticalSimilarityRepeating,cui2022LuminosityDistributionFast,Zhong2022} indicate that repeating FRBs and non-repeating FRBs differ in various properties, which may imply distinct physical mechanisms and origins. However, few previous studies are able to combine the differences in different properties and come up with a general case in distinguishing repeating and non-repeating FRBs. Moreover, because of the lack of continued observations for most FRBs, it is highly possible that some repeaters may have been incorrectly classified as non-repeaters. 

Machine learning (ML) can efficiently aid us in classifying FRBs with their traits and uncovering hidden repeaters. Machine learning is a branch of artificial intelligence (AI) and computer science dealing with complicated data sets that contain a large number of data points with numerous features. After setting a series of hyperparameters, machine learning algorithms can automatically train themselves and find the best parameters. In general, machine learning have been widely utilized in the detection of FRBs \citep[see e.g.][]{wagstaff2016MachineLearningClassifier,zhang2018FastRadioBursta,connor2018ApplyingDeepLearning,thechime/frbcollaboration2019ObservationsFastRadio,wu2019FeatureMatchingConditional,farah2019FiveNewRealtime,adamek2020SinglepulseDetectionAlgorithms,agarwal2020FETCHDeeplearningBased,yang202181NewCandidate}. 

Machine learning can be roughly divided into two types: supervised and unsupervised. Supervised machine learning takes target labels (classified results) as input data, while unsupervised ones do not require such a premise. Therefore, unsupervised machine learning is adept at revealing hidden information from the input data. 

In a series of two papers, we apply machine learning methods to perform classification of FRBs. The results using supervised methods have been reported in Paper I \citep{luo2022MachineLearningClassification}. In this paper, we report the results using unsupervised learning methods. 

Unsupervised machine learning is usually done in two steps. The first step is dimensionality reduction and the second step is clustering. Dimensionality reduction transforms and visualizes raw data from a high-dimensional space into a low-dimensional space so that we can easily see and evaluate it. Clustering is another kind of algorithm that clusters clumps of data points. We will discuss these steps in more detail in section \ref{subsec:dimensionality reduction} and section \ref{subsec:cluster}.

For FRB classification, \citet{Chen2022UMAP} used UMAP (one of the unsupervised machine learning algorithms) to identify hidden repeating FRBs. They presented repeating FRB candidates that are only captured once but the algorithm considered them as repeaters based on their similarities with the observed repeaters. But the small value of hyperparameter n\_neighbors used in this study may have caused the algorithm to be overly sensitive. \citet{2022Chaikova} built a pipeline to analyze the waveforms and concluded that there are two clusters. 

In this paper, we move a step further to consider more critical features of FRBs and utilize multiple unsupervised machine learning algorithms to analyze FRBs and cluster them based on their similarities in proprieties. We evaluate each algorithm and present our most credible repeater candidates for future repeater searches. In section \ref{sec:data}, we describe the selected input features of FRBs and other basic physical properties as input data, all from the first CHIME/FRB catalog \citep[][]{CHIME1}. In section \ref{sec:methods}, we introduce the workflows in our work, including three different and well-performed dimensionality reduction methods in section \ref{subsec:dimensionality reduction} and two corresponding cluster methods in section \ref{subsec:cluster}. The feature correlations are shown in section \ref{subsec:feature}. In section \ref{sec:analysis}, we analyze the performance of the above-mentioned algorithms, compare their visualized results, and reveal the physical difference between each cluster. We also discuss the number of FRB categories based on our unsupervised learning methods in section \ref{subsec:FRB num}. Then, we list our most plausible candidates for repeating FRBs that are consistent across all algorithms in Table \ref{tab: candidate}. In section \ref{sec:conclusion}, we make our conclusions based on results reported in previous sections. Comparing the candidates provided from Paper I \citep{luo2022MachineLearningClassification}, we identify a list of overlapping candidates and box them in Table \ref{tab: candidate}.

\section{Data}
\label{sec:data}

The input data are selected or derived from the first CHIME/FRB catalog \citep[][]{CHIME1} provided by the Canadian Hydrogen Intensity Mapping Experiment Fast Radio Burst (CHIME/FRB) Project in 2021. This catalog contains 536 FRBs, of which 474 are from non-repeating sources, and 62 are from 18 repeating FRB sources. To avoid massive impinge in taking the logarithm and in machine learning algorithms, we drop six FRBs with zero values for fluence (also for flux): FRB20190307A, FRB20190307B, FRB20190329B, FRB20190329C, FRB20190531A, and FRB20190531B. We also treat each sub-burst as an independent burst. Thus, we have 594 individual bursts including 500 from apparent non-repeaters and 94 from repeaters. 

To ensure the best performance of our machine learning algorithms, we choose the features that show different distributions between repeaters and non-repeaters. Among the observed properties that can be directly included from the CHIME catalog, we employ peak frequency, flux, fluence, and boxcar width. Other physical properties including redshift, rest-frame frequency width, rest-frame width, energy, luminosity, and brightness temperature, are derived based on the directly observed properties. We briefly describe these features below, and more detailed descriptions can be found in the companion Paper I \citep{luo2022MachineLearningClassification}.

The distributions of all the features are shown in Fig. \ref{fig:features}. We will delineate observed properties in section \ref{subsec:observed properties} and physical properties in section \ref{subsec:physical properties}.

\subsection{Observed Properties}
\label{subsec:observed properties}

Observed properties are the features directly selected from the first CHIME/FRB catalog. They show different distributions between repeaters and non-repeaters to some extent, which is also the foundation of machine learning.

\begin{itemize}
    \item Peak Frequency $\nu_c$ (MHz)
    
    Peak frequency is the sub-burst frequency of each FRB at its highest flux density. 
    
    \item Flux $S_{\nu}$ (Jy)
    
    Flux in the catalog is the Peak flux of the band-average profile (lower limit). Here we take their logarithmic values.
    
    \item Fluence $F_{\nu}$ (Jy ms)
    
    The fluence of FRBs in the first CHIME/FRB catalog is the flux integrated over time and we adopt their logarithmic values.
    
    \item Boxcar Width $\Delta t_{BC}$ (ms)
    
    In the first CHIME/FRB catalog, the burst duration is defined as the width of the boxcar after convolution \citep[][]{CHIME1}. We also take the logarithmic values.
    
    
\end{itemize}

\subsection{Physical Properties}
\label{subsec:physical properties}

Physical properties are derived from relatively complicated formulae that are linked more closely to the physical nature of FRBs. Compared with the observed properties, most of the physical properties differ more obviously between repeaters and non-repeaters. 

\begin{itemize}
    \item Redshift $z$
    
    For the two FRBs with photometric redshift measurements, FRB20121102A and FRB20180916B, we use their directly measured redshift values of $z=0.19273$ and $z=0.0337$ \citep{z_sp_Tendulkar_2017, z_sp_Marcote}. The redshifts of other FRBs are numerically solved from the dispersion measure (DM) of FRBs. 
    
    The observed DM of FRBs are consisted of four components \citep[e.g][]{Deng_2014_cos,Gao_2014_cos, z-DM_James}
    \begin{equation}
        {\rm DM=DM_{MW}+DM_{Halo}+DM_{IGM}}+\frac{\rm DM_{Host}}{1+z},
    \end{equation}
    corresponding to the contributions from the Milky Way disk, Milky Way halo, intergalactic medium, and FRB host galaxy. In this paper, we adopt NE2001 \citep{ne2001} to derive $\rm DM_{MW}$. 
    
    In the flat $\Lambda$CDM universe, $\rm DM_{IGM}$ can be written as \citep[][]{Deng_2014_cos,Zhang_2018_high_redshift,Hashimoto_2020,macquart2020CensusBaryonsUniverse}:
    \begin{equation}
        {\rm DM_{IGM}}(z) = \frac{3cH_0\Omega_b f_{\rm IGM}}{8\pi G m_p}\int_0^z \frac{\chi(z)(1+z)}{\sqrt{\Omega_m(1+z)^3+\Omega_{\Lambda}}}\,dz
    \end{equation}
    Here, $\chi(z)=Y_H X_{e,H}(z)+\frac{1}{2}Y_p X_{e,He}(z)$ is the fraction of ionized electrons to baryons in the IGM. We assume both H and He are fully ionized, i.e. $\chi(z)\sim 7/8$. Following \citet{1998Fukugita}, we adopt $f_{\rm IGM}=0.83$, which is the fraction of baryons in the IGM. For cosmological parameters, we take the latest Planck 18 results \citep[][]{Planck2018} for the $\Lambda CDM$ cosmology, $H_0=\SI{67.4}{\kilo\meter\per\second\per\mega\pc}$,$\Omega_b h^2=0.0224$,$\Omega_m=0.315$. As for $DM_{\rm halo}$ and $DM_{\rm host}$, we adopt $DM_{\rm halo}=\SI{30}{\pc.\cm^{-3}}$ and $DM_{\rm host}=\SI{70}{\pc.\cm^{-3}}$, following previous studies \citep[][]{DM_Yamasaki_2020, DM_Arcus, DM_Dolag, Hashimoto_2020, DM_Xu_2015}. 
    
    To avoid zero or negative values, we set a minimum redshift of 0.002248 corresponding to a luminosity distance of $\SI{10}{\mega\pc}$. 
    
\item Rest-frame Frequency Width $\Delta\nu$ (MHz)
    \begin{equation}
        \Delta \nu=(\nu_{max}-\nu_{min})(1+z)
    \end{equation}
    Rest-frame frequency width ($\Delta \nu$) is the difference between the highest frequency to the lowest frequency as observed, corrected for the cosmological redshift effect.
    
    \item Rest-frame Width $\Delta t_r$ (ms)
    
    \begin{equation}
        \Delta t_r=\frac{\Delta t}{1+z}
    \end{equation}
    Rest-frame width is the width of sub-burst using \texttt{fitburst} $\Delta t$ corrected for the time-dilation effect. We take their logarithmic values.
    
    \item Burst Energy $E$ (erg)
    
    Burst energy of FRBs can be deduced through \citep[][]{Zhang_2018_high_redshift}
    \begin{equation}
        E=\frac{4\pi D_L^2}{1+z}\mathcal{F}_\nu \nu_c,
    \end{equation}
    where $\mathcal{F}_\nu$ is the specific fluence, $D_L$ is the luminosity distance \citep[][]{cos_dis} and $\nu_c$ is the observed peak frequency of the FRB. We take their logarithmic values.
    
    \item Luminosity $L$(erg/s)
    
    The luminosity of FRBs can be derived from \citep[][]{Zhang_2018_high_redshift}:
    \begin{equation}
        L=4\pi D_L^2 \mathcal{S}_{\nu,p} \nu_c
    \end{equation}
    Where $\mathcal{S}_{\nu,p}$ is the specific peak flux and $\nu_c$ is the observed peak frequency of the FRB as above. We take their logarithmic values.
    
    \item Brightness Temperature $T_B$ (K)
    
    In radio astronomy, astronomers express the brightness of a source assuming the source is emitting a hypothetical blackbody with a temperature $T_B$  \citep[][]{radio_astronomy}. 
    With cosmological corrections, the brightness temperature in FRBs can be derived as (Paper I, \citet{luo2022MachineLearningClassification})
    \begin{equation}
        T_B=\frac{\mathcal{S}_{\nu,p}D_A^2}{2\pi k_B(\nu \Delta t)^2}(1+z)^3,
    \end{equation}
    where $k_B$ is the Boltzmann constant, $\nu$ is the peak frequency. 
    
\end{itemize}

\begin{figure*}
	\includegraphics[width=0.99\textwidth]{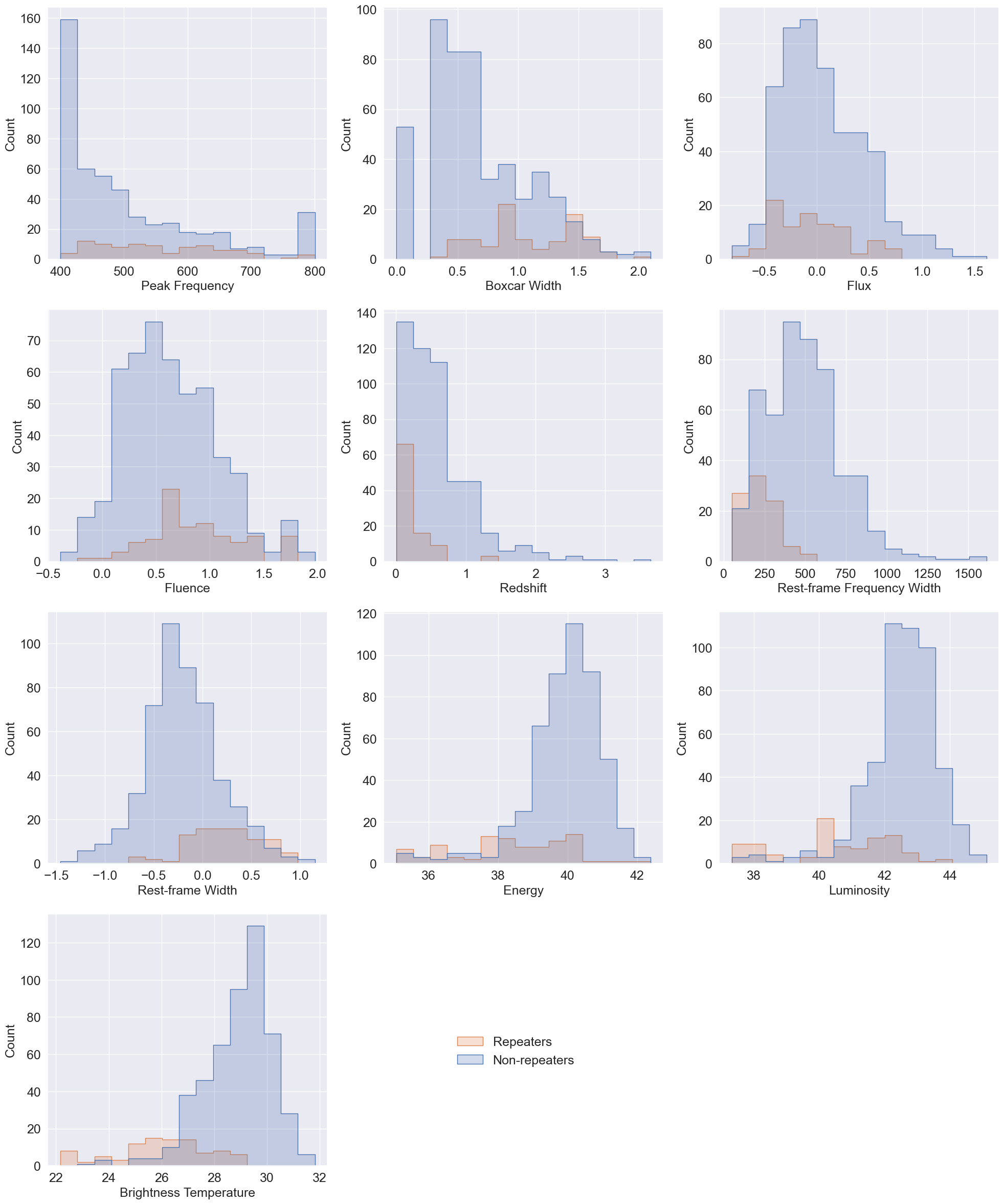}
	\caption{Comparison of the input features distributions between repeating and non-repeating FRBs.}
	\label{fig:features}
\end{figure*}

\section{Methods}
\label{sec:methods}

Two kinds of unsupervised machine learning methods, clustering and dimensionality reduction, are used in this paper. Dimensionality reduction algorithms learn high-dimensional data sets and automatically transform them into a low-dimensional space. Clustering algorithms group a set of data points into clusters based on their similarities. In practice, high-dimension data are usually first visualized by dimensionality reduction to a lower dimension. Then, clustering methods are used to identify clusters of the data points, after which we can label them manually. 

We utilize two kinds of dimensionality reduction algorithms, linear and manifold-based ones \citep[][]{manifold_Cayton2005}, in three different workflows. In each workflow, we first input the FRBs with 10 features derived in section \ref{sec:data} into dimensionality reduction methods to transfer them to the lower dimension. To better illustrate the results, we visualize them in a two-dimensional space. Therefore, for all dimensionality reduction algorithms used in this paper, we set the corresponding hyperparameter \texttt{n\_components} equal to 2. Note that in all the results of dimensional reduction shown in the figures, in order to compare with the cluster results, we color the repeaters in red and the non-repeaters in blue. However, we do not input the observed repetition of FRBs into the algorithms. Then, since the data points in the plot display some degree of clustering, we utilize clustering methods to identify them. Finally, to reveal the importance of the features in unsupervised machine learning, we deduce feature correlation from Mutual Information (MI). All the values of the hyperparameters used in this study are presented in Appendix \ref{sec: hyperparameter}. We list them in Table \ref{tab: dimension hyperparameters list} and Table \ref{tab: cluster hyperparameters list}, for dimensionality reduction and clustering respectively.


\subsection{Dimensionality Reduction}
\label{subsec:dimensionality reduction}

\subsubsection{Principal Component Analysis (PCA)}
\label{subsubsec:pca}

Principal Components Analysis (PCA) \citep[][]{PCA_Jolliffe1986, PCA_Hotelling1933AnalysisOA} is one of the most popular algorithms for linear dimensionality reduction. Given a data set, the PCA algorithm finds the directions (vectors) along which the data has a maximum variance and deduces the relative importance of these directions. Then, the algorithm keeps the most principal vectors as the principal components according to their importance. The number of reserved vectors is decided by the hyperparameter \texttt{n\_components}.

\begin{figure}
	\includegraphics[width=0.48\textwidth]{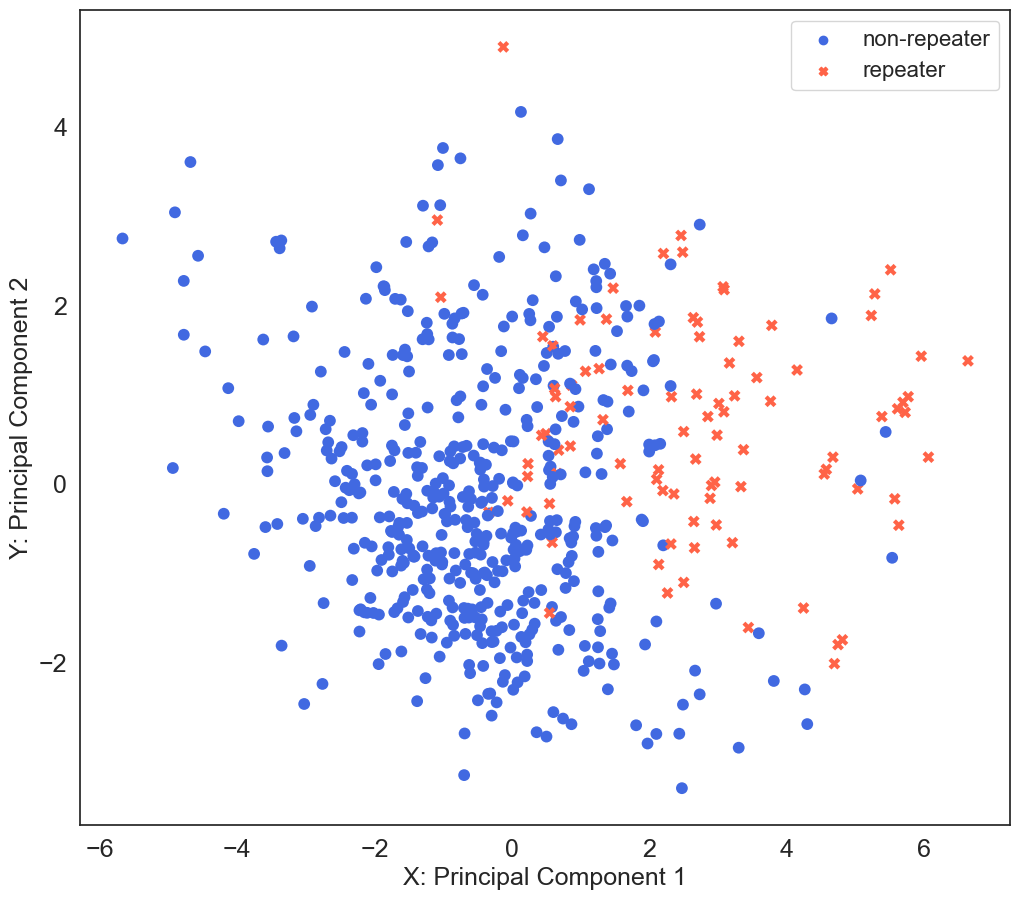}
	\caption{Dimensionality reduction results from PCA. The samples of repeating FRBs and non-repeating FRBs are clearly distributed in different regions.}
	\label{fig:PCA}
\end{figure}

Apart from \texttt{n\_components}, in the PCA algorithm, we adopt other hyperparameters in default values. Since PCA is a linear method based on variances, we preprocess the data to standardize features by removing the mean and scaling to unit variance and then putting them into the PCA algorithm. The dimensional reduction result of PCA is shown in Figure \ref{fig:PCA}. Repeaters and non-repeaters are clearly in different regions.

\subsubsection{t-distributed Stochastic Neighbor Embedding (t-SNE)}
\label{subsubsec:tsne}

T-distributed Stochastic Neighbor Embedding (t-SNE) \citep[][]{tsne_visual,tsne_acc} is a manifold dimensionality reduction algorithm. It is based on Stochastic Neighbor Embedding (SNE) \citep[][]{sne}. Stochastic Neighbor Embedding (SNE) first converts the high-dimensional Euclidean distances into conditional probabilities. Then, SNE algorithms minimize the sum of Kullback-Leibler divergences by optimizing the cost function with gradient descent. However, SNE is prone to the ``crowding problem'' which is common in manifold algorithms and can show crowded point distributions in the two-dimensional space. In this case, t-SNE modifies the Gaussian distribution used in the probability of SNE to a Student-t distribution. Compared with Gaussian distribution, t distribution has a heavy tail, causing two distant data points to increase their distance while transforming to a lower dimension. Therefore, t-SNE can avoid the crowding problem and its performance is greatly improved. 

\begin{figure}
	\includegraphics[width=0.48\textwidth]{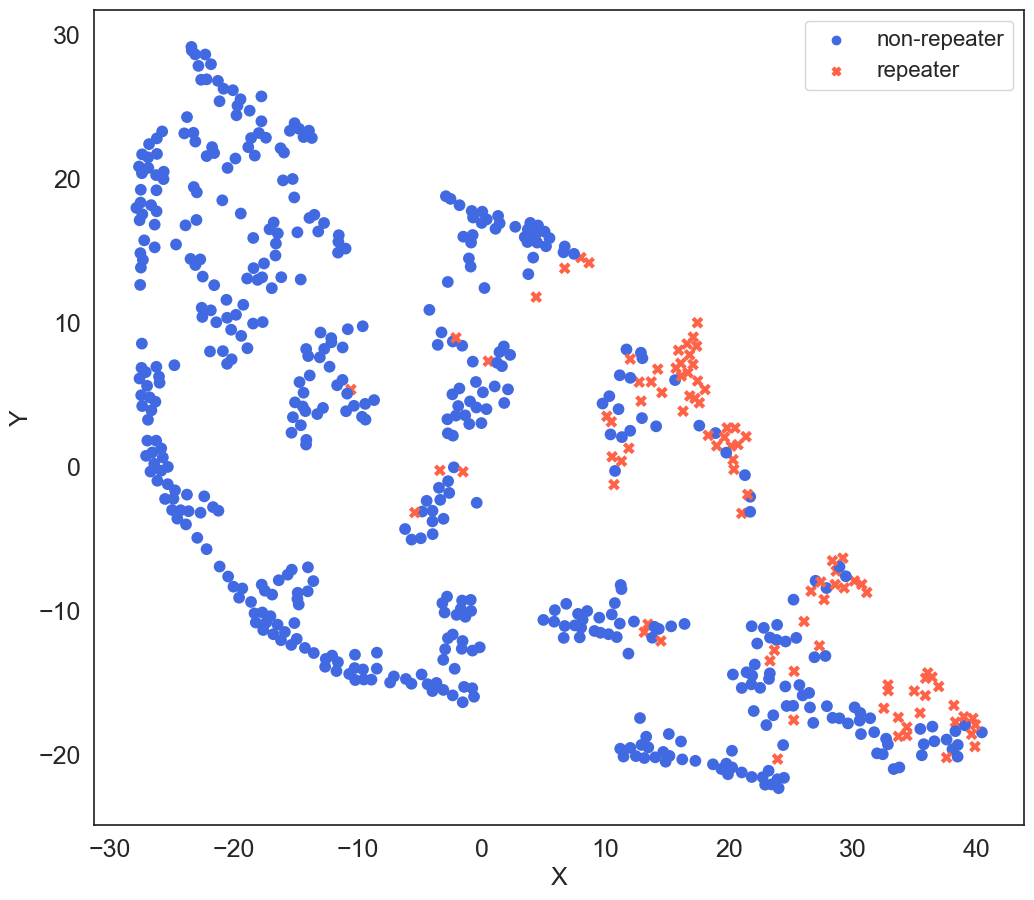}
	\caption{Dimensionality reduction results from t-SNE. Most repeating FRBs are on the right-hand side and generally reside in the same clusters.}
	\label{fig:TSNE}
\end{figure}

Perplexity is one of the critical hyperparameters in t-SNE. It is related to the number of nearest neighbors. A large value of perplexity means a more global view of the manifold, while a small value shows the local structure of the input. According to \cite{oskolkov_2019_hyperparameter}, $\mathrm{perplexity}\sim N^{1/2}$ is a reasonable value, while $N$ is the number of input samples. Thus, we set perplexity equal to 24. Another hyperparameter \texttt{n\_iter} stands for the number of iterations in the optimization as mentioned before. In gradient descent, if the number of iterations is large enough, the parameters will be closer to their local optimal solutions but cost much time. Here we set \texttt{n\_iter} as 1000, which is enough number for the iterations. The dimensionality reduction results are plotted in Figure \ref{fig:TSNE}. Most repeating FRBs are on the right-hand side and generally reside in the same clusters.

\subsubsection{Uniform Manifold Approximation and Projection (UMAP)}
\label{subsubsec:umap}

Uniform Manifold Approximation and Projection (UMAP) \citep[][]{umap} is another manifold learning technique for dimensionality reduction. As a manifold algorithm, the theoretical foundation for UMAP is also based on Riemannian geometry. However, UMAP assumes that the input data is uniformly distributed on the manifold. Another difference in UMAP is the modification from metric spaces to fuzzy topological representations. 

\begin{figure}
	\includegraphics[width=0.48\textwidth]{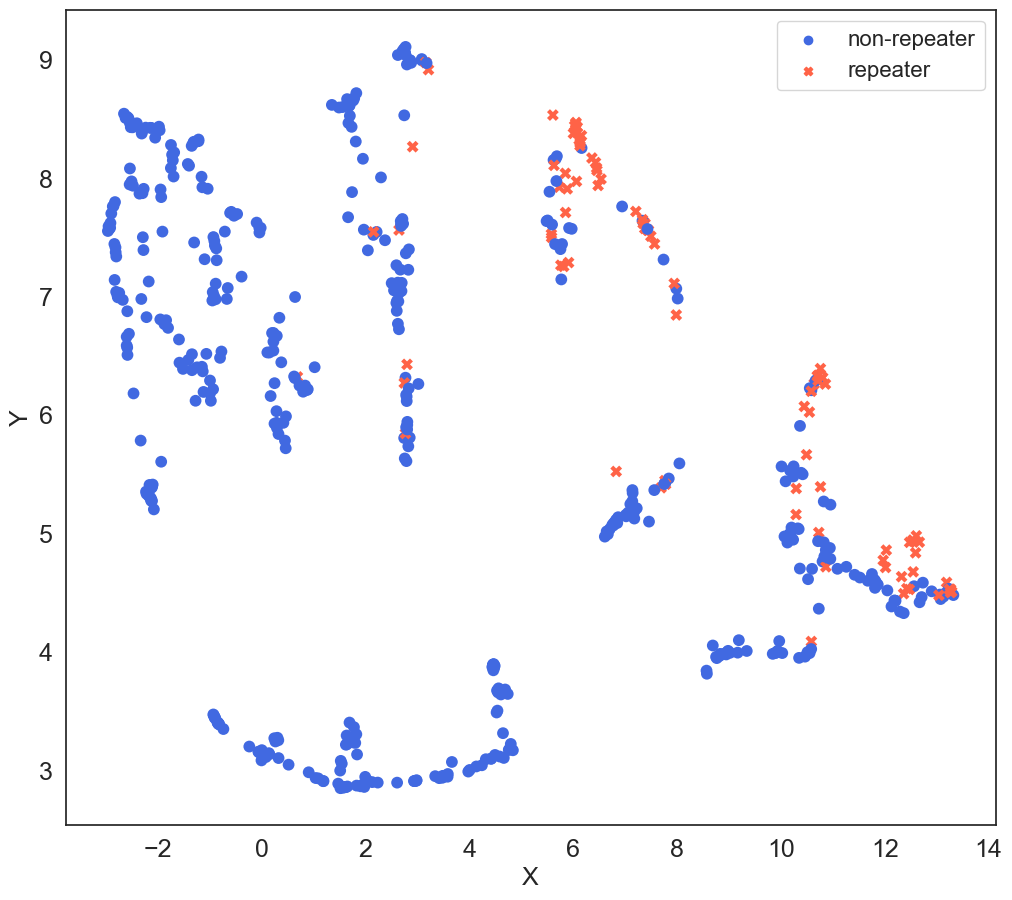}
	\caption{Dimensionality reduction results from UMAP. Similar to t-SNE, most repeating FRBs are on the right-hand side and generally reside in the same clusters.}
	\label{fig:UMAP}
\end{figure}

Two hyperparameters \texttt{n\_neighbors} and \texttt{min\_dist} count here. Similar to perplexity in t-SNE, \texttt{n\_neighbors} corresponds to the size of the local neighborhood used for manifold approximation. A higher value of \texttt{n\_neighbors} results in a more global view of the manifold. Similarly, we adopt \texttt{n\_neighbors} to be 24. Another vital hyperparameter \texttt{min\_dist} represents the effective minimum distance among embedded points. To recognize various clusters without interference by the internal points in each cluster, for UMAP, a small value should be set. Therefore, we adopt \texttt{min\_dist} to be 0 based on our experience. UMAP's results are shown in Figure \ref{fig:UMAP}. Similar to t-SNE, most repeating FRBs are on the right-hand side and generally reside in the same clusters.

\subsection{Clustering}
\label{subsec:cluster}

\subsubsection{$k$-means}
\label{subsubsec:k-means}

The $k$-means algorithm \citep[][]{k-means_Lloyd, k-means_MacQueen1967SomeMF} groups the data points into clusters by minimizing the sum of squares of Euclidean distances between the geometric points and their centroids. It first initializes $k$ points as cluster centers and then optimizes their positions until they reach the real centers of each cluster. Because $k$-means is based on the distance from each point to their centers, it performs well in circular-like clusters but fails to identify clusters with strange shapes such as curved shapes. Therefore, we use $k$-means in dealing with linear-based dimensionality reduction algorithms PCA. 

\begin{figure}
	\includegraphics[width=0.48\textwidth]{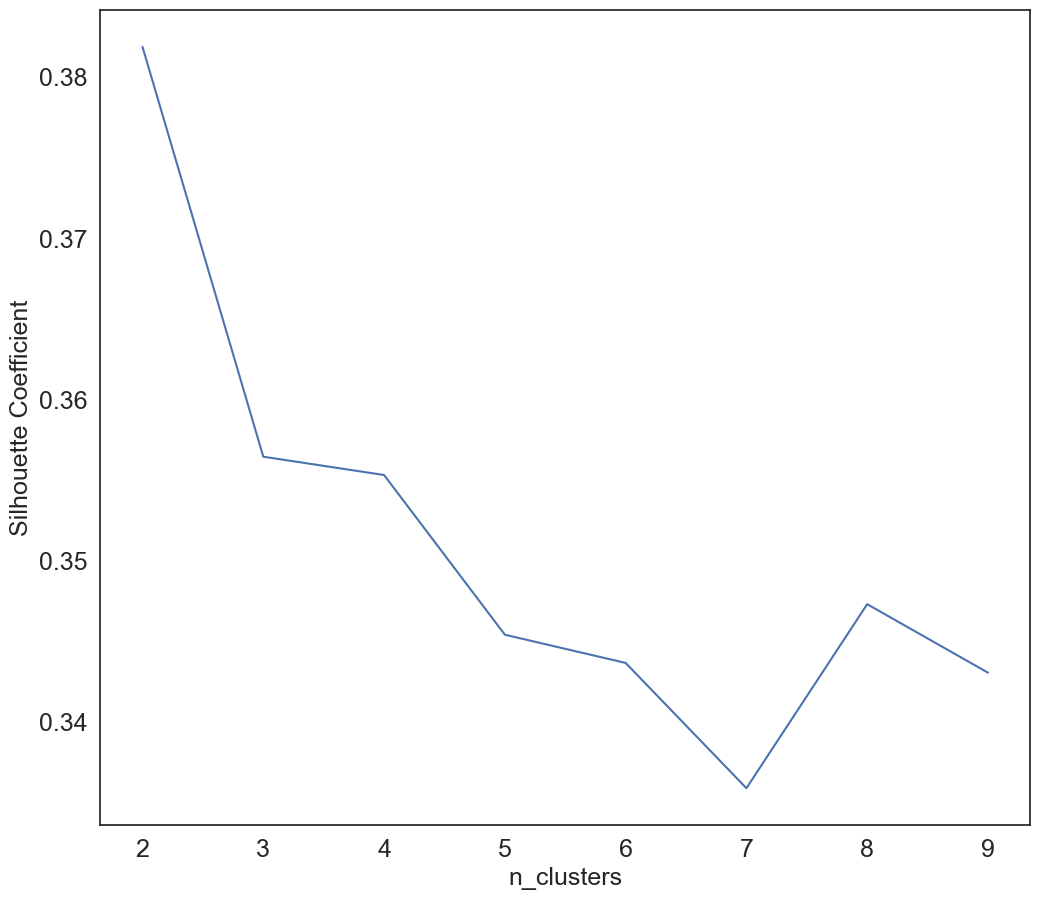}
	\caption{Silhouette coefficients of $k$-means after PCA with respect to different number of clusters. The maximum value of silhouette coefficient implies 2 clusters in PCA space and the corresponding hyperparameter for $k$-means.}
	\label{fig:PCA K-means Coefficient}
\end{figure}

\begin{figure}
	\includegraphics[width=0.48\textwidth]{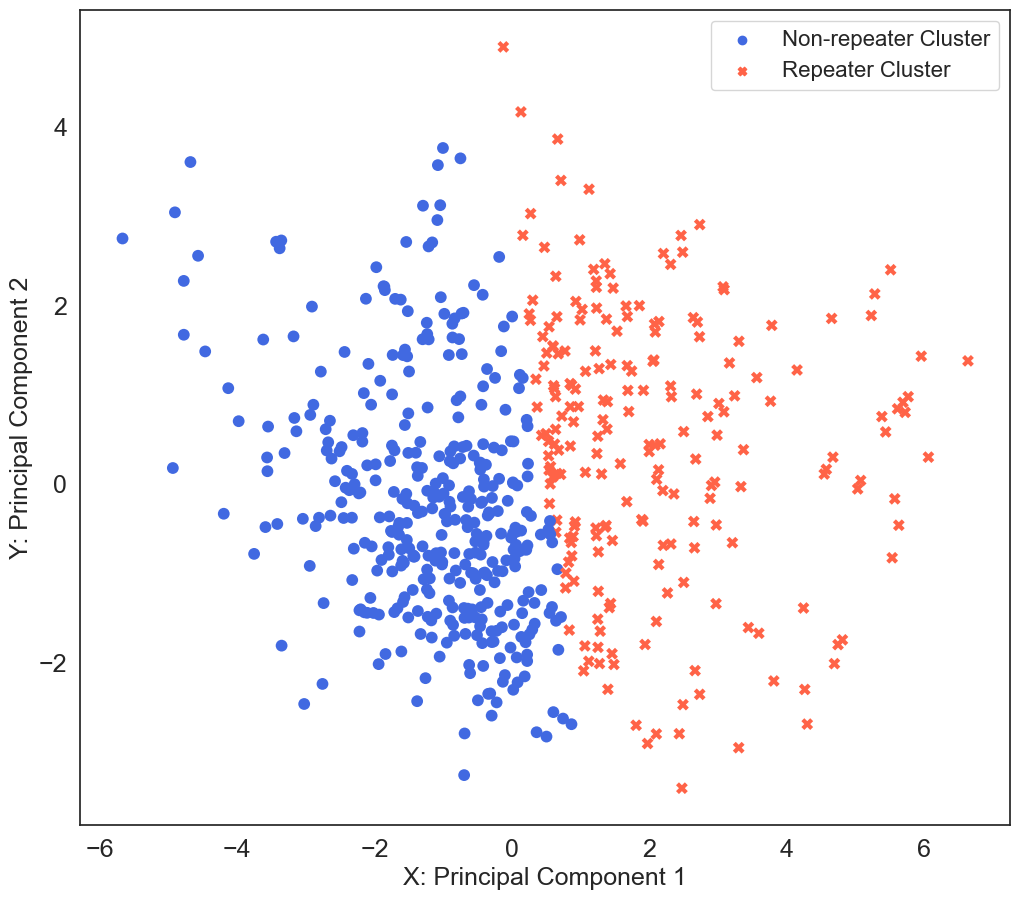}
	\caption{The $k$-means clustering result in PCA space. The right cluster, containing a higher ratio of repeaters, is identified as the repeater cluster, while the left one is identified as the non-repeater cluster. The two clusters are consistent with the distribution of observed repeaters and non-repeaters in Figure \ref{fig:PCA}.}
	\label{fig:PCA_Kmeans}
\end{figure}

A vital hyperparameter in $k$-means is \texttt{n\_clusters}, which refers to how many cluster centers are present in the model. In this paper, we calculate the silhouette coefficient of $k$-means with different \texttt{n\_clusters} to determine the best number of clusters. Silhouette coefficient has long been used to evaluate the quality of clustering. It ranges from -1 to 1, and a higher value stands for more coherent clusters. The silhouette coefficient for one of the data points $o_i$ is defined as \citep[][]{Silhouettes_1987_Rousseeuw, Silhouettes_2012_Jiawei, Silhouettes_2021_Hoss}:
\begin{equation}
    s(o_i)=\frac{b(o_i)-a(o_i)}{max\{a(o_i),b(o_i)\}},
\end{equation}
where $a(o_i)$ is the average distance between $o_i$ and all other data points within the same cluster, while $b(o_i)$ refers to the average distance between $o_i$ and all other data points outside the cluster to which $o_i$ belongs. For the clustering metric, the silhouette coefficient is the average value for all data points. If a clustering method performs well, each cluster should be tightly clustered and have a long distance from other clusters. In this case, its silhouette coefficient will be close to 1. The values of silhouette coefficients in various \texttt{n\_clusters} of $k$-means are shown in Figure \ref{fig:PCA K-means Coefficient}. We choose \texttt{n\_clusters} equal to 2 corresponding to the maximum value of the silhouette coefficient and show the cluster result in Figure \ref{fig:PCA_Kmeans}.  


\subsubsection{Hierarchical Density-Based Spatial Clustering of Applications with Noise (HDBSCAN)}
\label{subsubsec:hdbscan}

Another algorithm we use to cluster the data points is Hierarchical Density-Based Spatial Clustering of Applications with Noise (HDBSCAN)\citep[][]{hdbscan2,hdbscan1,McInnes2017}. HDBSCAN is an algorithm improved from Density-Based Spatial Clustering of Applications with Noise (DBSCAN) \citep[][]{dbscan_Ester1996} and combined with Hierarchical Clustering \citep[][]{Silhouettes_2012_Jiawei, Silhouettes_2021_Hoss}. DBSCAN utlizes two hyper-parameters to estimate the density in an area: \texttt{$\varepsilon$} and \texttt{MinPts}. \texttt{$\varepsilon$} is the radius defined as the neighborhood for a point, and \texttt{MinPts} is defined as the critical number of data points within the neighborhood. Three types of points are then defined: 
\begin{enumerate}
    \item Core point: a point that has more data points within its neighborhood than \texttt{MinPts}.
    \item Border point: a point in the neighborhood of one core point but the number of data points within its neighborhood is fewer than \texttt{MinPts}.
    \item Noise point: a point that is neither a core point nor a border point.
\end{enumerate}
With specific values of \texttt{$\varepsilon$} and \texttt{MinPts}, DBSCAN automatically clusters the data points based on the area covering the core points and border points without any parameter corresponding to the number of clusters. Hierarchical clustering is a series of methods that start from the cluster of each point and merge to a larger cluster, or from a single cluster and split to smaller subclusters, according to the change of the criterion value. HDBSCAN is refined from DBSCAN in simplifying the density measurement and combining with hierarchical clustering while scanning the densities. It chooses the clusters based on the hierarchical structure from such steps. 

HDBSCAN performs well after manifold dimensionality reduction over $k$-means due to the strange shapes of the groups. Those groups usually are not in circles or follow Gaussian distribution. $k$-means or other clustering algorithms fail to apply. Therefore, we utilize HDBSCAN after the manifold algorithms t-SNE and UMAP. Similar to DBSCAN, HDBSCAN does not need to know the number of clusters in advance. HDBSCAN automatically determines the number of clusters while clustering the points. Besides, HDBSCAN can recognize outlier points as noises. 

We first input data into the t-SNE or UMAP algorithms to reduce the dimension and then utilize the HDBSCAN algorithm to label the clusters. The clusters that only include non-repeaters are recognized as "non-repeater clusters". The clusters containing more than a single-digit number (we defined 15\% as the criterion) of repeaters are classified as "repeater clusters". Non-repeaters in repeater clusters can then be treated as hidden repeaters. As for the clusters that only have a few repeaters and the ratio is not enough for 15\%, they are labeled as "other clusters". The single-digit repeaters in these clusters may be misclassified due to the inaccuracies of their input features, or indicate that the clusters are repeater ones. Further observation is required. 

\begin{figure}
	\includegraphics[width=0.48\textwidth]{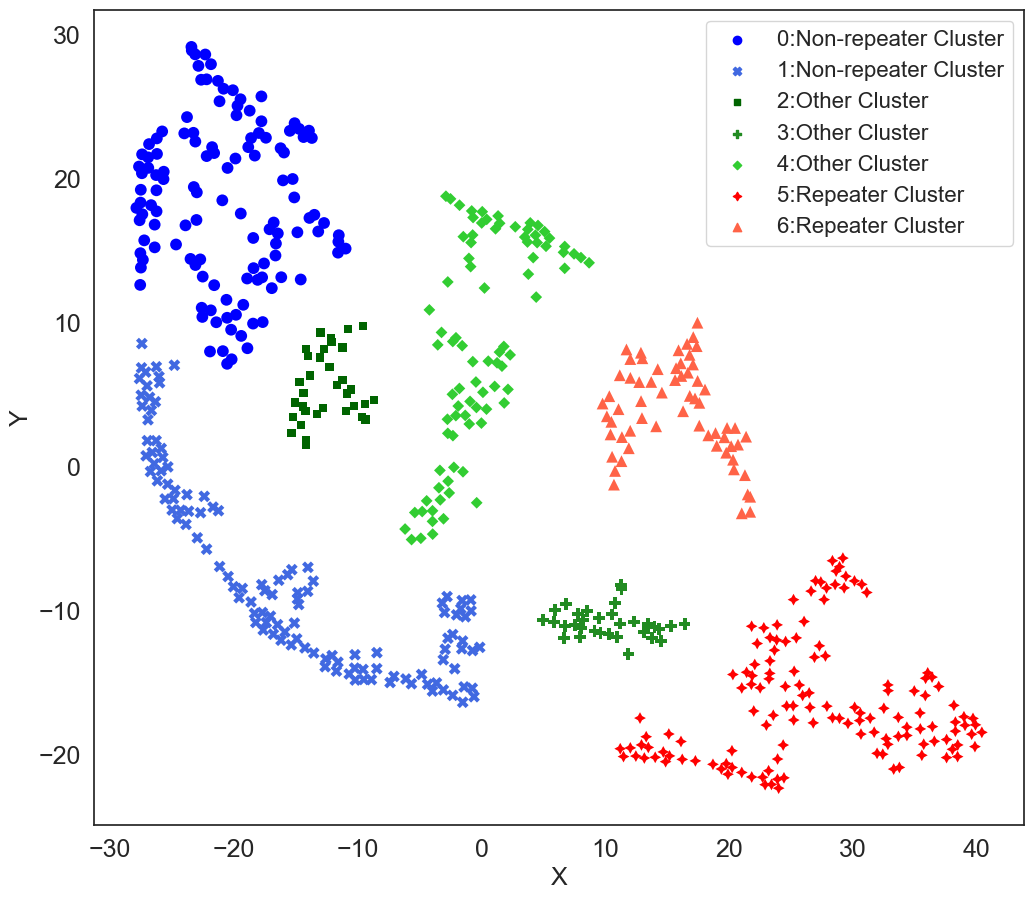}
	\caption{Clustering results of HDBSCAN in the t-SNE plane. Non-repeater clusters are marked in blue, while repeater clusters are marked in red. Other clusters are marked in green.}
	\label{fig:TSNE_HD}
\end{figure}

\begin{figure}
	\includegraphics[width=0.48\textwidth]{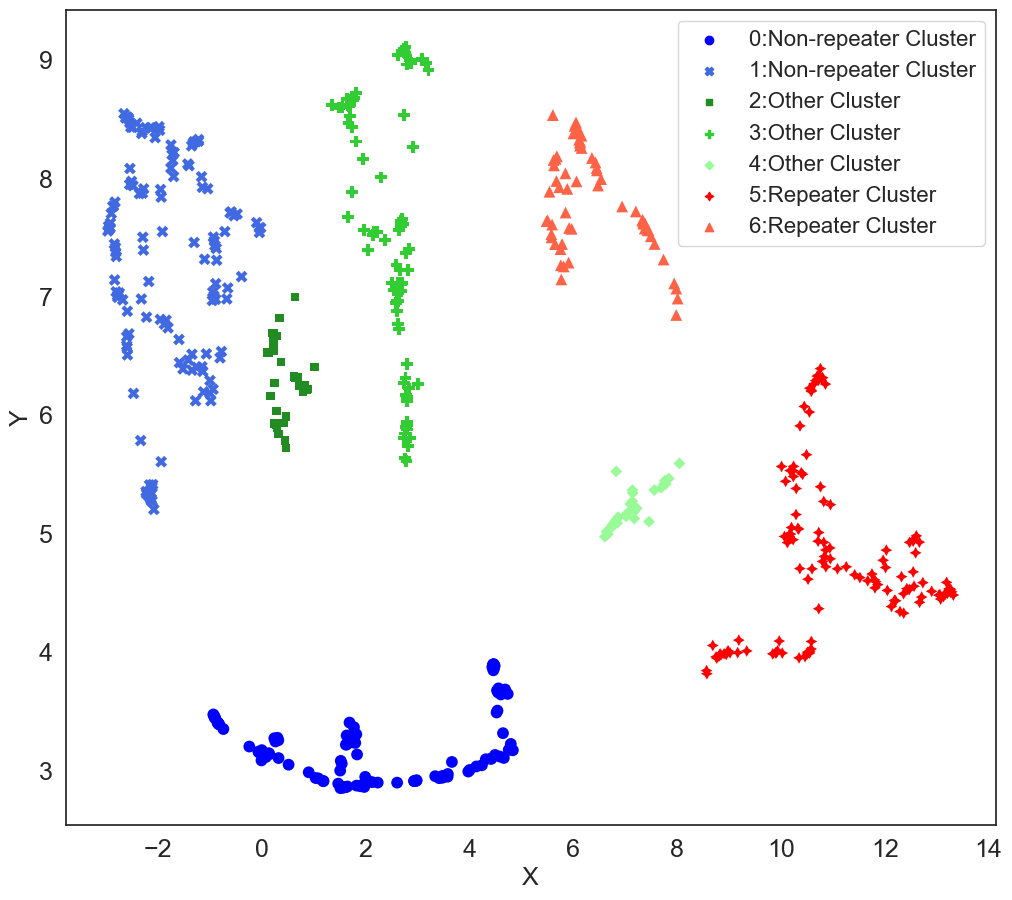}
	\caption{Clustering results of HDBSCAN in the UMAP plane. Non-repeater clusters are marked in blue, while repeater clusters are marked in red. Other clusters are marked in green.}
	\label{fig:UMAP_HD}
\end{figure}

The cluster results of t-SNE and UMAP are shown respectively in Figure \ref{fig:TSNE_HD} and Figure \ref{fig:UMAP_HD}. To better visualize them, we color the repeater clusters in red color series, other clusters in green color series, and non-repeater clusters in blue color series.

\subsection{Feature Correlations: Mutual Information}
\label{subsec:feature}




After dimensionality reduction and clustering, the data points are visualized in two-dimensional spaces and are clustered based on their similarities. However, the correlation between the final result and the input features is still unknown. In order to unravel the correlation and the importance of features in different algorithms, Mutual Information regression in the \texttt{scikit-learn} library is used to estimate the Mutual Information between the features and the coordinates \citep[][]{MI_flow}. 

Mutual Information is first proposed by \citet{MI_Shannon} and is widely adopted in machine learning since \cite{MI_battiti}. MI is calculated from the Kullback–Leibler divergence between the joint distribution and the product of the marginal distributions of two variables. For a pair of random variables $X$ and $Y$, MI can be written as:
\begin{equation}
    I(X;Y)=\int \int P_{(X,Y)}(x,y)\log \frac{P_{(X,Y)}(x,y)}{P_{X}(x)P_Y(y)}\,dx\,dy,
\end{equation}
where $P_{(X,Y)}$ is the joint probability mass function of $X$ and $Y$, and $P_{X}$ and $P_{Y}$ are the marginal probability mass functions of $X$ and $Y$, respectively. It is a non-negative value and measures the mutual dependence between the two variables, as a higher MI score implies a higher dependence.

\begin{figure}
	\includegraphics[width=0.48\textwidth]{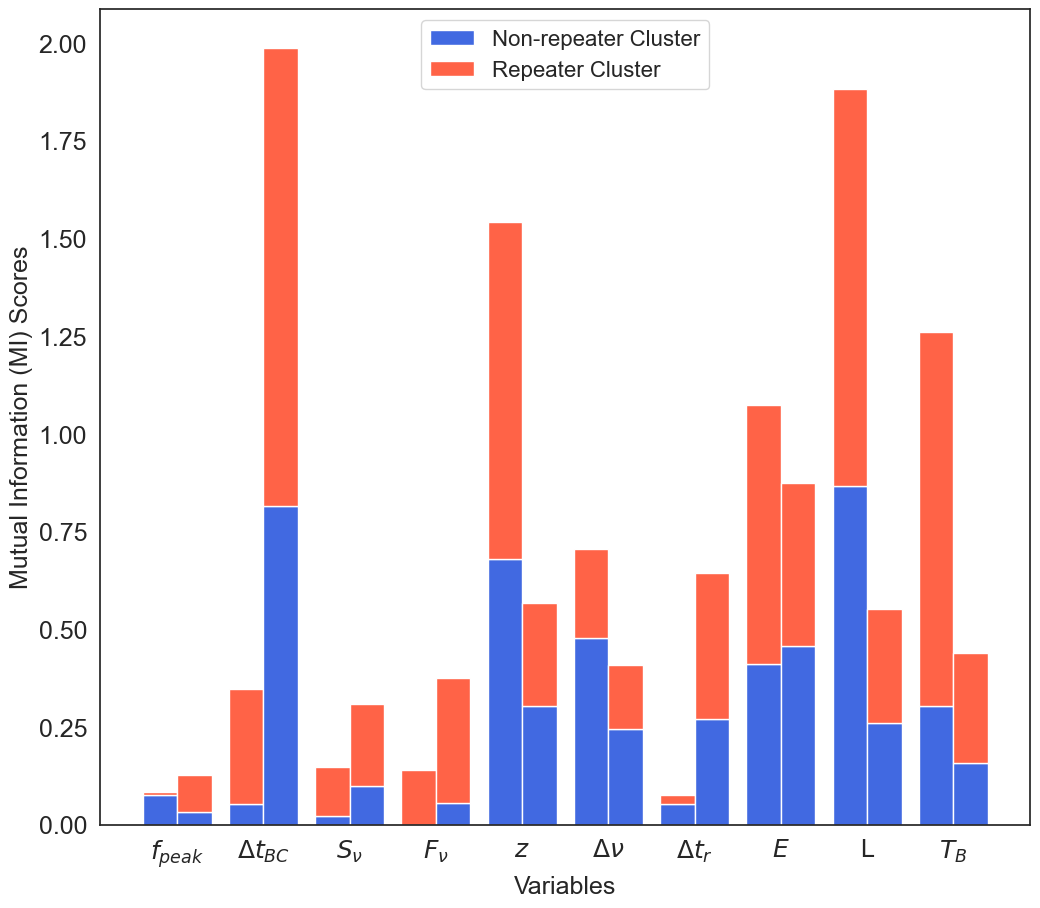}
	\caption{Feature correlation of PCA+$k$-means. The left bar correlates with the $x$-axis while the right bar is related to the $y$-axis. Boxcar width, redshift, energy, luminosity and brightness temperature are the most important features in PCA.}
	\label{fig:PCA_features}
\end{figure}

\begin{figure}
	\includegraphics[width=0.48\textwidth]{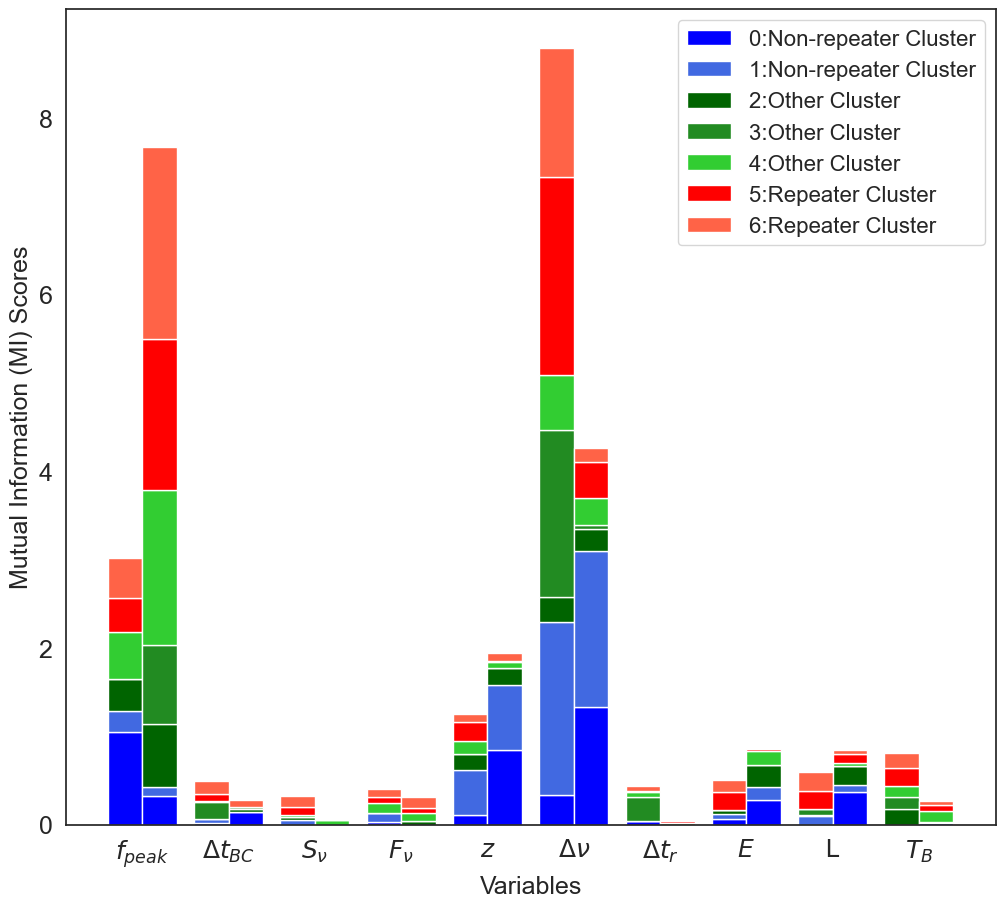}
	\caption{Feature correlation of t-SNE+HDBSCAN. For non-repeater clusters in blue, peak frequency, redshift and rest-frame frequency width are the most dominant features, while redshift is not as significant in the repeater clusters colored red. On the whole, peak frequency, redshift and rest-frame frequency width are the most important features. }
	\label{fig:TSNE_features}
\end{figure}

\begin{figure}
	\includegraphics[width=0.48\textwidth]{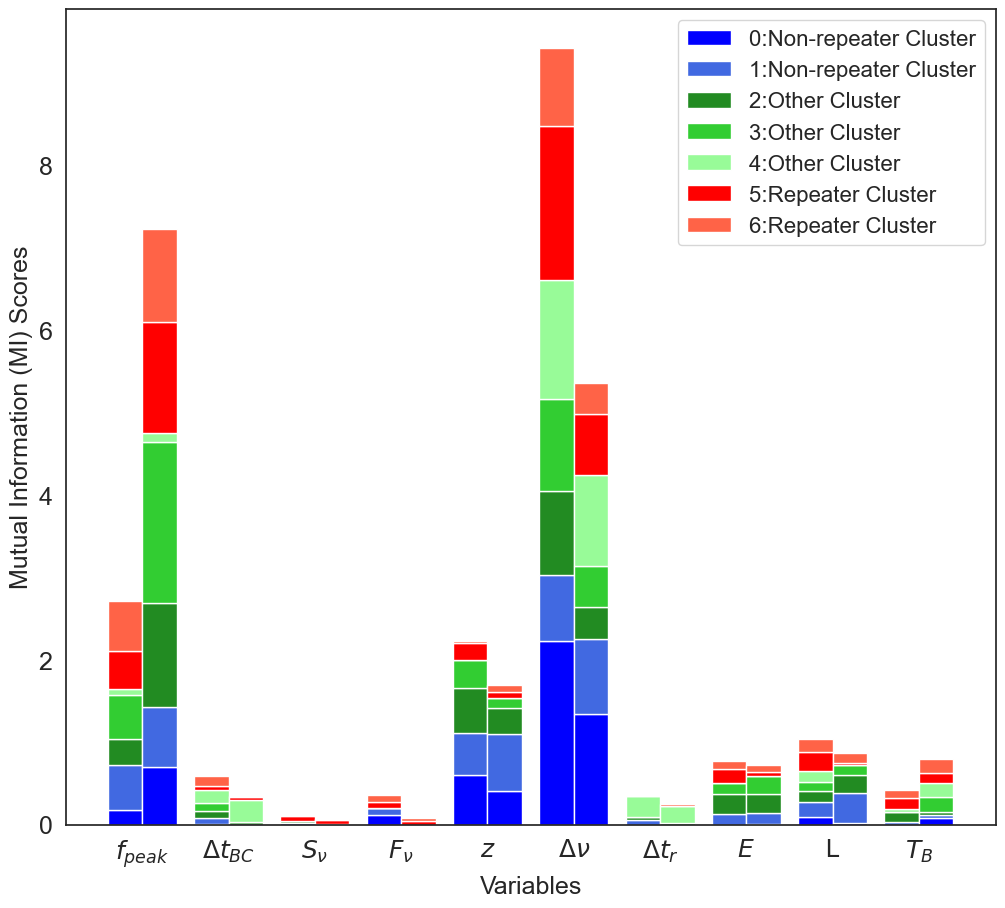}
	\caption{Features correlation of UMAP+HDBSCAN. For non-repeater clusters in blue, peak frequency, redshift and rest-frame frequency width are the most dominant features, while redshift is not as significant in the repeater clusters colored red. On the whole, peak frequency, redshift and rest-frame frequency width are the most important features.}
	\label{fig:UMAP_features}
\end{figure}

In this paper, we first compute the MI scores of each cluster to estimate the feature correlation for the clusters. Then, we stack the MI scores of the clusters in each method to show the importance of features in dimensionality reduction. To better illustrate them, we paint the bars in the same color series as clustering results reported in section \ref{subsec:cluster}. Note that for each feature, the left bar correlates with the $x$-axis while the right bar is related to the $y$-axis. The MI scores of PCA, t-SNE and UMAP are shown in Figures \ref{fig:PCA_features}, \ref{fig:TSNE_features} and \ref{fig:UMAP_features}.

In Figure \ref{fig:PCA_features}, the MI scores of the linear method PCA are relatively balanced between the repeater cluster and non-repeater cluster, which is consistent with the clustering results in Figure \ref{fig:PCA_Kmeans}. Generally, boxcar width, redshift, energy, luminosity and brightness temperature are the most important features in PCA. The MI scores also reflect the variances of features in the principal vectors.

For the manifold methods t-SNE and UMAP, clustered by HDBSCAN, their feature correlation shown in Figure \ref{fig:TSNE_features} and \ref{fig:UMAP_features} are similar. For non-repeater clusters in blue, peak frequency, redshift and rest-frame frequency width are the most dominant features, while redshift is not as significant in the repeater clusters colored red. On the whole, peak frequency, redshift and rest-frame frequency width are the most important features.

\section{Analysis and Discussion}
\label{sec:analysis}

In section \ref{sec:methods}, we presented the models we selected and their visualized results. Since they are all unsupervised machine learning methods, no information on burst identification (whether an FRB is a  repeater or an apparent non-repeater) is included. The algorithms only study their features and transform them into different clusters. Then, by comparing with the observed identities from the CHIME catalog, we consider various clusters as repeaters, non-repeaters, or others. In this section, we will first introduce some metrics to evaluate the performance of machine learning algorithms. Then, we delineate the similarities between manifold methods. After that, we combine all the results, deduce their weight vector to evaluate their significance, and rank each FRB to obtain its classified label. Finally, we list our most credible repeater candidates.

\subsection{Evaluation of different models}
\label{subsec:evaluation}

In machine learning classification, outputs are written as the following forms:
\begin{enumerate}
    \item True positive (TP): 
    
    Predicted to be positive, and is truly positive.
    
    \item True negative (TN): 
    
    Predicted to be negative, and is truly negative.
    
    \item False positive (FP): 
    
    Predicted to be positive, but is actually negative.
    
    \item False negative (FN): 
    
    Predicted to be negative, but is actually positive.
\end{enumerate}
In our work, we define the positive as repeaters, while the negative as other than repeaters (including non-repeaters clusters and other clusters). Also, under the definition, repeater candidates are False Positives (FPs). 

Based on the definition of outcomes and our classification results, we can get a general sense that if a model covers almost all the actual repeaters in its predicted repeaters and includes a few repeater candidates (FP here), this model can be considered well-performing. Precision and recall are two metrics that meet our needs.
\begin{itemize}
    \item Precision
    \begin{equation}
        P=Precision=\frac{TP}{TP+FP}
    \end{equation}
    \item Recall
    \begin{equation}
        R=Recall=\frac{TP}{TP+FN}
    \end{equation}
\end{itemize}
Notice that we do not adopt accuracy here, which is defined as $(TP+TN)/(TP+FN+FP+TN)$. This is because we need a model with a high ratio of TP, and a low ratio of FN and FP, rather than the ratio of TN. Even if each model categorizes all the points as non-repeaters, $Accuracy$ will be highly scored.

To combine both precision and recall, we adopt F-measure to evaluate the models. The full definition of the F-measure is given by \citep[][]{M4_metrics_Chinchor,metrics_IR_David}:
\begin{equation}
    F_\beta = \frac{(1+\beta^2)PR}{\beta^2 P+R}\qquad (0\leq \beta \leq +\infty).
\end{equation}
Here, values of $\beta$ represent weights of $precision$ and $recall$ \citep[][]{metrics_F_Sasaki}. Specially, if $\beta=1$, $F_\beta$ falls back to the classic $F_1-score$, where $precision$ and $recall$ are considered equally. Apart from that, if $\beta < 1$, $F_\beta$ becomes more dominated by $precision$ and if $\beta >1$, $F_\beta$ are more recall-oriented. For FRBs, apparently non-repeaters may actually be repeaters, while observed repeaters should not be classified as non-repeaters. In this case, $Recall$ must be a priority and preferably to be close to 100\%. Therefore, We adopt $\beta=2$ and calculate the metrics in Table \ref{tab:results}. All workflows show $F_2-score$ higher than 0.70, indicating good performances.

\begin{table*}
	\caption{Evaluation metrics for each model. The high scores in $F_2$ indicate good performances of the three workflows.}
	\label{tab:results}
	\begin{tabular}{c|cccc|cc|c}
		\hline
		Method & TP & FN(Misclassified Repeaters) & FP(Repeaters Candidates) & TN & $Recall$ & $Precision$ & $F_2$ \\
		\hline
        PCA+$k$-means & 85 & 9 & 127 & 373 & 0.9042 & 0.4009  & 0.7227  \\
        t-SNE+HDBSCAN & 81 & 13  & 117 & 383 & 0.8617 & 0.4091  & 0.7056  \\
        UMAP+HDBSCAN & 81 & 13  & 117 & 383 & 0.8617 & 0.4041  & 0.7056\\
        \hline
	\end{tabular}
\end{table*}

\subsection{Manifold Structures across Methods}
\label{subsec:manifold structure across methods}

\begin{figure}
	\includegraphics[width=0.48\textwidth]{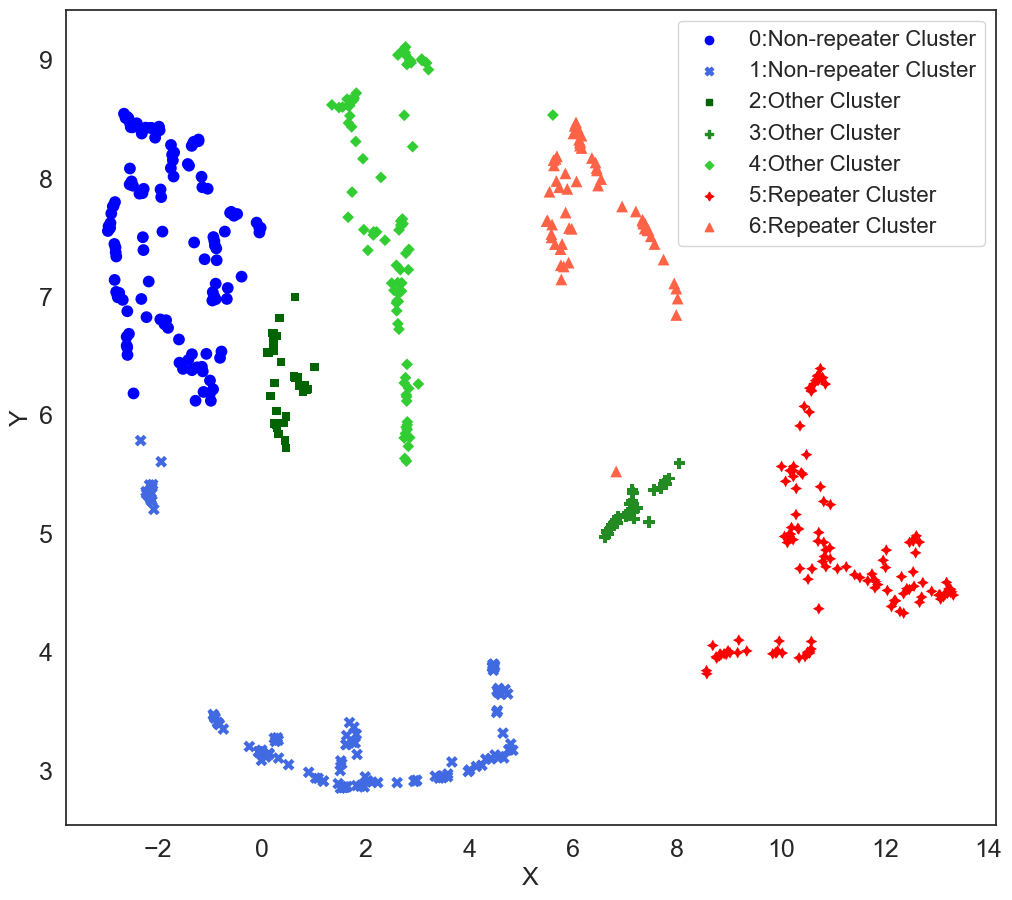}
	\caption{Data points in UMAP plane colored with the cluster labels identified by t-SNE and HDBSCAN. Only two points are in the different series of clusters.}
	\label{fig:TSNE_HD_UMAP}
\end{figure}

\begin{figure}
	\includegraphics[width=0.48\textwidth]{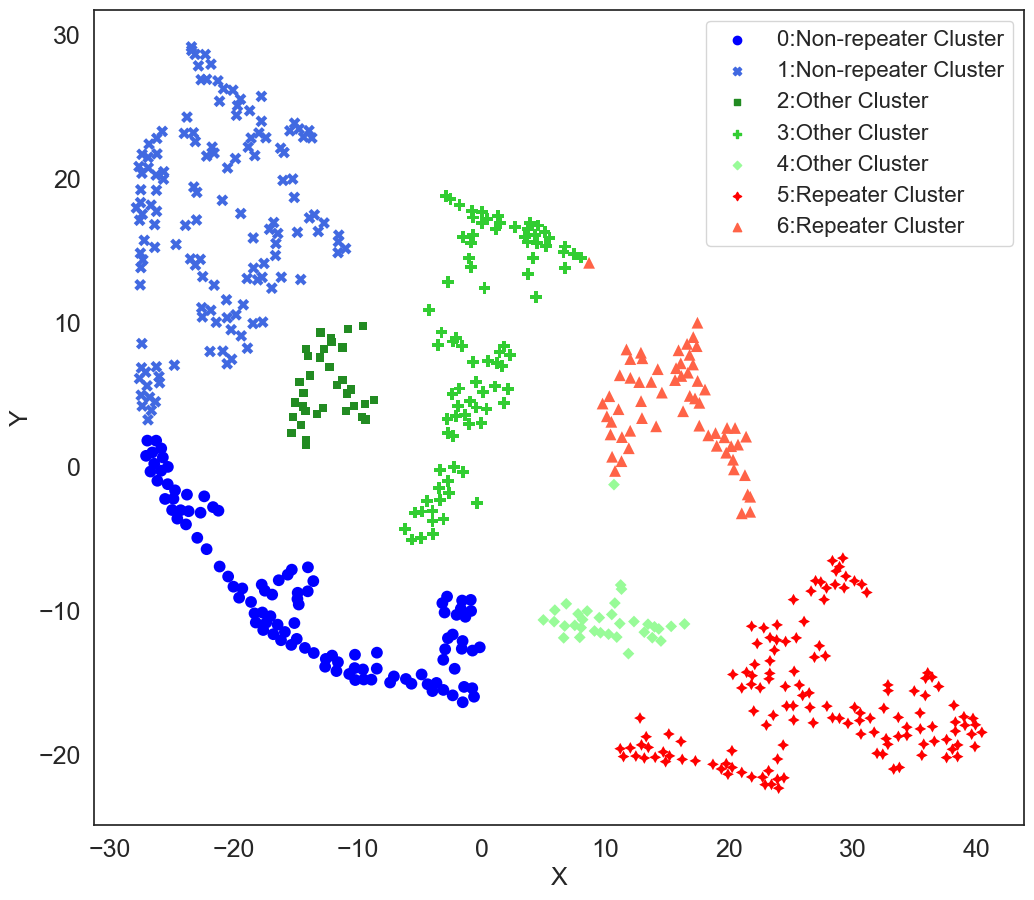}
	\caption{Data points in t-SNE plane colored with the cluster labels identified by UMAP and HDBSCAN. Only two points are in the different series of clusters.}
	\label{fig:UMAP_HD_TSNE}
\end{figure}

Until now, the analysis for the results and evaluations are all given individually. But since we employ different algorithms and visualize them, an interesting practice is to study the structures across different methods. Especially, manifold results in Figure \ref{fig:TSNE_HD} and Figure \ref{fig:UMAP_HD} seem to share some groups with similar shapes. A simple idea is to plot one classification set on another reduced dimensional plane to check if those clusters are the same FRBs. Here we plot the t-SNE result in UMAP space in Figure \ref{fig:TSNE_HD_UMAP} and the exchange condition in Figure \ref{fig:UMAP_HD_TSNE}. In both figures, the color series across the mehods are generally the same, only a few points on the edge of clusters may be in different clusters. 

\subsection{Categories of FRBs}
\label{subsec:FRB num}

Since the physical origin of FRBs is not identified, we do not know how many categories of FRBs are there.
Unsupervised machine learning algorithms learn the data from their features without any premise of repeaters and non-repeaters, those algorithms open a window to uncover the number of categories and also the traits of each category. 

\begin{figure}
	\includegraphics[width=0.48\textwidth]{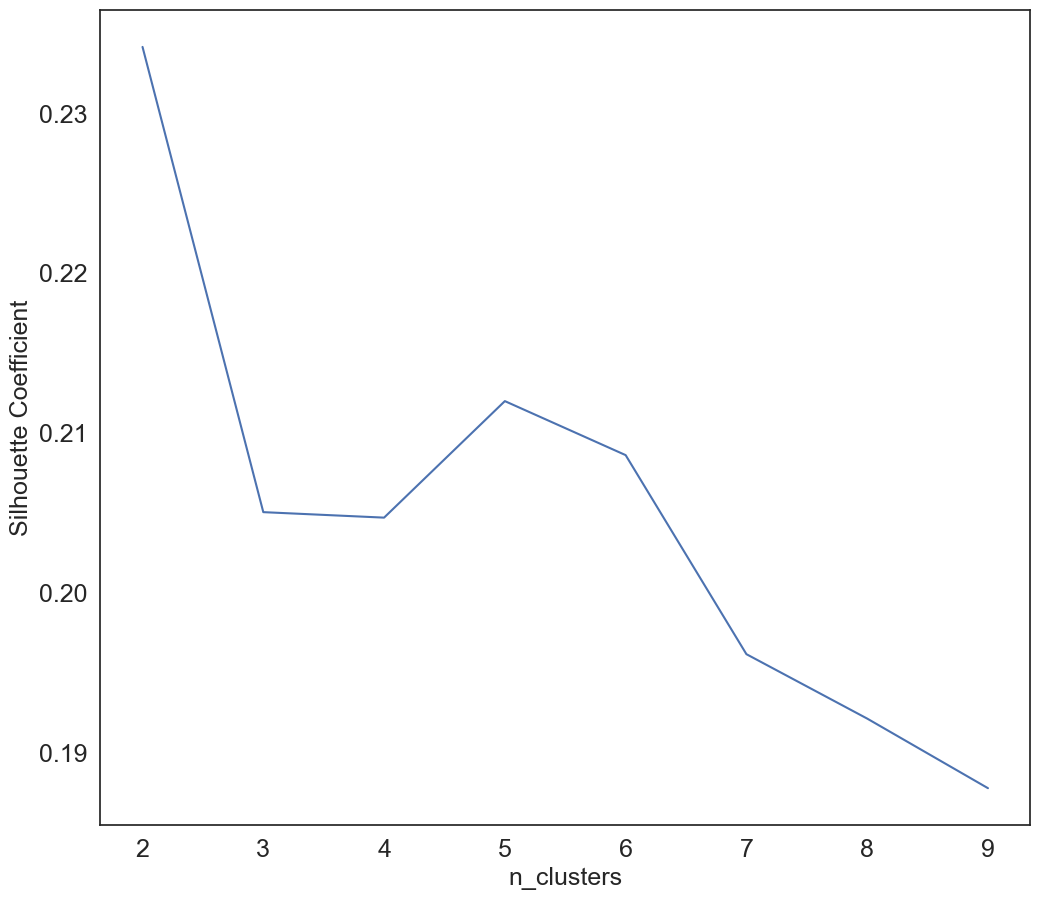}
	\caption{Silhouette coefficients of $k$-means with respect to different number of clusters, directly in the raw space without PCA. In addition to the maximum value at 2 clusters, the plateau corresponding to 5 and 6 clusters implies possible subcategories with lower significance. }
	\label{fig:Silhouette Coefficient raw}
\end{figure}

In section \ref{subsubsec:k-means}, we calculated silhouette coefficient to determine the number of clusters. Here, we can also deduce the coefficients directly in the raw high-dimensional space. The result shown in Figure \ref{fig:Silhouette Coefficient raw} indicates 2 main categories of FRBs, while 5 or 6 is also possible albeit with lower significance. 

\begin{figure*}
	\includegraphics[width=0.99\textwidth]{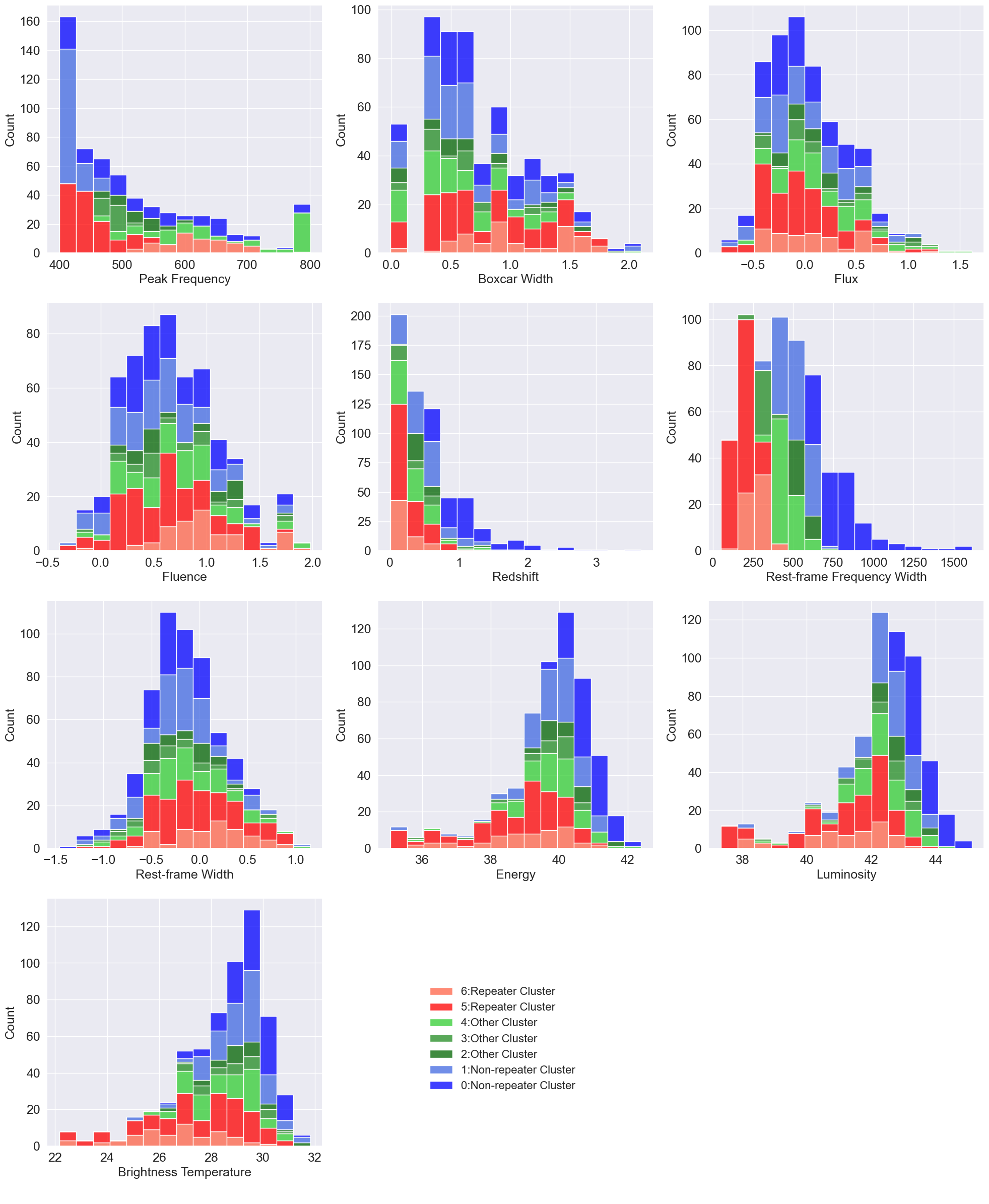}
	\caption{Input features of FRBs in the labels given by t-SNE and HDBSCAN. The different distributions of the clusters imply the corresponding categories. }
	\label{fig:feature_label}
\end{figure*}

Another clue is from the manifold methods in section \ref{subsubsec:tsne} and section \ref{subsubsec:umap}, both algorithms tend to output 7 categories in the two-dimensional space. In order to reveal the features of each cluster more specifically, considering the similar classified results of manifold methods discussed in section \ref{subsec:manifold structure across methods}, we plot the features again but in hues of the labels given by t-SNE in Figure \ref{fig:feature_label}. Note that peak frequency and rest-frame frequency width of most FRBs in clusters 1 and 5 are at the observational limit of CHIME, which might be two false sub-categories because of observational biases, indicating that there might be 5 to 7 species of FRBs. Compared with other research applied unsupervised machine learning to Gamma-ray Bursts \citep{2020_GRB_ML_Christian}, which just includes two clear categories, the number of species in FRBs might imply a more complicated physical origin than GRBs.

In Figure \ref{fig:feature_label}, the repeater clusters share smaller values in redshift, rest-frame frequency width, energy, luminosity and brightness temperature, and slightly larger values in boxcar width, which is consistent with the observed distributions of repeaters in Figure \ref{fig:features}. Specifically, in the non-repeater clusters, cluster 0 is the non-repeater cluster with relatively higher peak frequency, redshift, rest-frame frequency width, energy and luminosity. For the repeater clusters, cluster 6 is the repeater cluster with higher peak frequency, fluence and rest-frame frequency width. For the other clusters, their properties lie between the repeater clusters and non-repeater clusters. While the non-repeaters in the repeater clusters are those repeater candidates we want, those in the other clusters also need follow-up observations to identify their real classes.

\subsection{Repeater candidates from the Combination of Algorithms}
\label{subsec:combination}

We used various algorithms to classify FRBs in the first CHIME/FRB catalog. Since they all provide a set of hidden repeaters, their intersection can be reckoned as the most credible repeaters candidates. Thus, we give the candidate list, hoping to assist future repeater search. Their basic information and features are listed in Table \ref{tab: candidate}. We also cross compare with the repeater candidates identified using supervised methods (Paper I, \citet{luo2022MachineLearningClassification}) and box the overlapping candidates from the two papers. These should be the strongest candidates and we encourage close follow-up observations of these sources to detect their repeated bursts.

\begin{table*}
\caption{Repeater candidates: the most plausible candidates given by unsupervised machine learning methods. The boxed FRBs are also identified as repeater candidates in supervised methods (Paper I, \citet{luo2022MachineLearningClassification}). Sub num is the sub-burst number of FRBs in the first CHIME/FRB catalog. Description of the properties can be found in Section \ref{sec:data}.
}
\label{tab: candidate}
\addtolength{\tabcolsep}{-1pt}
\scalebox{0.90}{
\begin{tabular}{llllllllllllll}
\hline
\multirow{2}{*}{Name} & Sub & RA & Dec & $\nu_c$ & $\log\Delta t_{BC}$ & $\log S_{\nu}$ & $\log F_{\nu}$ & $z$ & 
$\Delta\nu$ & $\log\Delta t_r$ & $\log E$ & $\log L$ & $\log T_B$ \\
& Num & $^\circ$ & $^\circ$ & $\si{\mega\hertz}$ & $\log\si{\milli\s}$ & $\log\si{\Jy}$ & $\log\si{\Jy\milli\s}$ & & $\si{\mega\hertz}$ & $\log\si{\milli\s}$ & $\log\si{\erg}$ & $\log\si{\erg\per\s}$ & $\log\si{\kelvin}$\\
\hline
FRB20180907E & 0   & 167.88 & 47.09 & 400.2             & 1.07    & -0.14 & 0.84    & 0.312 & 178.54         & 0.500         & 39.85     & 41.99     & 27.89         \\
FRB20180911A & 0   & 99.55  & 84.62 & 400.2             & 0.29    & 0.20  & 0.41    & 0.085 & 216.59         & -0.116        & 38.26     & 41.09     & 28.62         \\
FRB20180915B & 0   & 225.23 & 25.02 & 400.2             & 0.69    & 0.00  & 0.58    & 0.072 & 136.33         & 0.199         & 38.28     & 40.73     & 27.47         \\
FRB20180920B & 0   & 191.09 & 63.52 & 421.1             & 1.03    & -0.46 & 0.23    & 0.401 & 116.53         & 0.221         & 39.49     & 41.95     & 27.83         \\
FRB20180923A & 0   & 327.61 & 71.92 & 468.9             & 0.29    & -0.12 & 0.08    & 0.026 & 177.17         & -0.835        & 36.96     & 39.77     & 27.12         \\
FRB20180923C & 0   & 239.14 & 22.85 & 420.5             & 0.29    & -0.05 & 0.14    & 0.059 & 291.70         & -0.520        & 37.69     & 40.53     & 28.01         \\
FRB20180928A & 0   & 312.95 & 30.85 & 400.2             & 0.47    & 0.13  & 0.40    & 0.002 & 92.11          & -0.571        & 35.08     & 37.81     & 25.02         \\
FRB20181013E & 0   & 307.28 & 69.02 & 400.2             & 0.47    & -0.21 & 0.31    & 0.208 & 269.06         & -0.147        & 38.96     & 41.52     & 28.66         \\
\boxed{FRB20181017B} & 0   & 237.76 & 78.50 & 593.2             & 1.11    & 0.03  & 0.81    & 0.207 & 247.97         & 0.282         & 39.63     & 41.92     & 27.27         \\
\boxed{FRB20181030E} & 0   & 135.67 & 8.89  & 470.5             & 0.77    & 0.30  & 0.80    & 0.013 & 166.76         & -0.404        & 37.09     & 39.59     & 25.99         \\
FRB20181125A & 0   & 147.94 & 33.93 & 434.5             & 1.17    & -0.41 & 0.51    & 0.171 & 156.33         & 0.039         & 39.02     & 41.17     & 26.81         \\
FRB20181125A & 1   & 147.94 & 33.93 & 436.6             & 1.17    & -0.41 & 0.51    & 0.171 & 177.76         & 0.090         & 39.02     & 41.17     & 26.81         \\
FRB20181125A & 2   & 147.94 & 33.93 & 426.5             & 1.17    & -0.41 & 0.51    & 0.171 & 141.34         & 0.130         & 39.01     & 41.16     & 26.83         \\
FRB20181130A & 0   & 355.19 & 46.49 & 412.1             & -0.01   & -0.01 & 0.10    & 0.033 & 119.55         & -0.333        & 37.14     & 40.04     & 28.16         \\
FRB20181214A & 0   & 70.00  & 43.07 & 435.0             & 0.47    & -0.81 & -0.39   & 0.231 & 116.19         & -0.363        & 38.39     & 41.06     & 28.08         \\
FRB20181220A & 0   & 346.11 & 48.43 & 400.2             & 0.47    & 0.12  & 0.48    & 0.002 & 196.64         & -0.362        & 35.16     & 37.81     & 25.02         \\
\boxed{FRB20181221A} & 0   & 230.58 & 25.86 & 510.1             & 0.69    & 0.10  & 0.76    & 0.240 & 170.15         & -0.216        & 39.65     & 42.07     & 28.44         \\
FRB20181226E & 0   & 303.56 & 73.64 & 400.2             & 0.47    & -0.32 & 0.13    & 0.178 & 186.23         & -0.003        & 38.64     & 41.26     & 28.41         \\
\boxed{FRB20181229B} & 0   & 238.37 & 19.78 & 445.5             & 1.31    & -0.38 & 0.69    & 0.320 & 154.80         & 0.406         & 39.77     & 41.83     & 27.09         \\
\boxed{FRB20181231B} & 0   & 128.77 & 55.99 & 657.7             & 0.47    & -0.05 & 0.37    & 0.066 & 276.56         & -0.500        & 38.22     & 40.82     & 27.37         \\
FRB20190106B & 0   & 335.63 & 46.13 & 452.1             & 0.29    & 0.23  & 0.58    & 0.098 & 157.27         & -0.278        & 38.61     & 41.30     & 28.67         \\
\boxed{FRB20190109B} & 0   & 253.47 & 1.25  & 408.1             & 0.84    & 0.08  & 0.48    & 0.009 & 93.05          & -0.473        & 36.40     & 39.00     & 25.45         \\
FRB20190110C & 0   & 246.98 & 41.42 & 427.4             & 0.47    & -0.19 & 0.15    & 0.112 & 86.17          & -0.455        & 38.27     & 40.98     & 28.06         \\
FRB20190111A & 2   & 217.00 & 26.78 & 400.2             & 0.77    & 0.56  & 1.23    & 0.066 & 251.25         & -0.130        & 38.87     & 41.22     & 27.81         \\
\boxed{FRB20190112A} & 0   & 257.98 & 61.20 & 697.7             & 0.99    & 0.15  & 1.21    & 0.348 & 317.48         & 0.085         & 40.56     & 42.63     & 27.94         \\
\boxed{FRB20190129A} & 0   & 45.06  & 21.42 & 707.7             & 0.95    & -0.31 & 0.70    & 0.404 & 347.23         & -0.094        & 40.19     & 42.33     & 27.70         \\
FRB20190204A & 0   & 161.33 & 61.53 & 418.2             & 0.90    & -0.62 & 0.18    & 0.382 & 217.88         & 0.117         & 39.39     & 41.73     & 27.90         \\
\boxed{FRB20190206A} & 0   & 244.85 & 9.36  & 534.5             & 0.77    & 0.15  & 0.96    & 0.062 & 213.84         & -0.121        & 38.66     & 40.87     & 27.08         \\
\boxed{FRB20190218B} & 0   & 268.70 & 17.93 & 588.0             & 1.25    & -0.24 & 0.77    & 0.442 & 334.17         & 0.153         & 40.26     & 42.41     & 27.41         \\
FRB20190220A & 0   & 237.21 & 74.16 & 400.2             & 0.29    & -0.47 & -0.17   & 0.098 & 265.91         & -0.300        & 37.81     & 40.55     & 28.08         \\
FRB20190221A & 0   & 132.60 & 9.90  & 444.8             & 0.47    & 0.09  & 0.37    & 0.092 & 225.37         & -0.052        & 38.34     & 41.10     & 28.14         \\
FRB20190222B & 0   & 160.69 & 19.62 & 400.2             & 1.34    & -0.40 & 0.64    & 0.439 & 278.25         & 0.024         & 39.96     & 42.08     & 27.41         \\
FRB20190223A & 0   & 64.72  & 87.65 & 444.8             & 0.59    & -0.33 & 0.20    & 0.286 & 149.61         & -0.227        & 39.18     & 41.76     & 28.48         \\
FRB20190228A & 0   & 183.48 & 22.90 & 664.7             & 1.48    & 0.25  & 1.55    & 0.365 & 357.33         & 0.217         & 40.93     & 42.76     & 27.15         \\
FRB20190308C & 0   & 188.36 & 44.39 & 453.4             & 1.34    & -0.33 & 0.68    & 0.454 & 218.85         & -0.561        & 40.09     & 42.24     & 27.40         \\
FRB20190308C & 1   & 188.36 & 44.39 & 449.0             & 1.34    & -0.33 & 0.68    & 0.454 & 211.58         & -0.422        & 40.08     & 42.24     & 27.41         \\
FRB20190308B & 0   & 38.59  & 83.62 & 477.2             & 0.29    & 0.05  & 0.14    & 0.015 & 190.59         & -0.737        & 36.57     & 39.48     & 26.81         \\
FRB20190308B & 1   & 38.59  & 83.62 & 455.5             & 0.29    & 0.05  & 0.14    & 0.015 & 187.95         & -0.291        & 36.55     & 39.46     & 26.85         \\
FRB20190323D & 0   & 56.88  & 46.93 & 400.2             & 1.11    & -0.43 & 0.40    & 0.593 & 204.86         & 0.533         & 39.99     & 42.36     & 28.10         \\
\boxed{FRB20190329A} & 0   & 65.54  & 73.63 & 432.3             & 1.07    & -0.28 & 0.35    & 0.002 & 73.87          & 0.016         & 35.06     & 37.43     & 23.34         \\
FRB20190403E & 0   & 220.22 & 86.54 & 620.6             & 1.27    & 0.59  & 1.88    & 0.099 & 348.03         & 0.301         & 40.06     & 41.81     & 26.81         \\
\boxed{FRB20190409B} & 0   & 126.65 & 63.47 & 545.5             & 1.48    & -0.41 & 0.83    & 0.175 & 336.98         & 0.299         & 39.46     & 41.29     & 26.01         \\
\boxed{FRB20190410A} & 0   & 263.47 & -2.37 & 515.7             & 0.84    & 0.20  & 0.76    & 0.073 & 182.92         & -0.026        & 38.59     & 41.06     & 27.18         \\
\boxed{FRB20190412B} & 0   & 285.65 & 19.25 & 400.2             & 1.63    & -0.17 & 1.11    & 0.015 & 228.59         & 0.826         & 37.42     & 39.15     & 24.04         \\
FRB20190418A & 0   & 65.79  & 16.04 & 400.2             & 0.29    & 0.00  & 0.34    & 0.019 & 182.96         & -0.159        & 36.89     & 39.55     & 27.11         \\
FRB20190419A & 0   & 104.98 & 64.88 & 407.1             & 0.59    & -0.39 & -0.11   & 0.341 & 185.53         & 0.140         & 38.99     & 41.84     & 28.66         \\
FRB20190422A & 0   & 48.56  & 35.15 & 626.1             & 1.47    & -0.22 & 0.96    & 0.335 & 373.37         & 0.382         & 40.23     & 42.18     & 26.68         \\
\boxed{FRB20190422A} & 1   & 48.56  & 35.15 & 612.3             & 1.47    & -0.22 & 0.96    & 0.335 & 312.63         & 0.238         & 40.22     & 42.17     & 26.70         \\
FRB20190423A & 1   & 179.68 & 55.25 & 400.2             & 0.77    & 1.03  & 1.74    & 0.143 & 220.29         & 0.316         & 40.06     & 42.40     & 28.96         \\
\boxed{FRB20190423B} & 0   & 298.58 & 26.19 & 537.6             & 0.99    & -0.06 & 0.85    & 0.003 & 159.79         & 0.395         & 35.93     & 38.03     & 23.81         \\
\boxed{FRB20190423B} & 1   & 298.58 & 26.19 & 524.6             & 0.99    & -0.06 & 0.85    & 0.003 & 148.96         & 0.928         & 35.92     & 38.02     & 23.83         \\
\boxed{FRB20190429B} & 0   & 329.93 & 3.96  & 422.4             & 1.22    & -0.13 & 0.70    & 0.194 & 50.64          & 0.728         & 39.31     & 41.56     & 27.12         \\
FRB20190430A & 0   & 77.70  & 87.01 & 433.8             & 1.29    & -0.12 & 0.89    & 0.228 & 214.13         & 0.440         & 39.65     & 41.73     & 27.10         \\
FRB20190517C & 0   & 87.50  & 26.62 & 435.5             & 0.29    & 0.49  & 0.94    & 0.063 & 147.92         & -0.450        & 38.56     & 41.14     & 28.57         \\
\boxed{FRB20190527A} & 0   & 12.45  & 7.99  & 484.7             & 1.76    & -0.33 & 1.00    & 0.537 & 205.46         & 0.240         & 40.59     & 42.44     & 26.65         \\
FRB20190527A & 1   & 12.45  & 7.99  & 449.1             & 1.76    & -0.33 & 1.00    & 0.537 & 172.11         & 0.206         & 40.55     & 42.41     & 26.72         \\
FRB20190531C & 0   & 331.14 & 43.00 & 453.0             & 0.47    & -0.43 & 0.08    & 0.304 & 183.03         & 0.046         & 39.12     & 41.73     & 28.67         \\
FRB20190601B & 0   & 17.88  & 23.82 & 429.5             & 1.39    & 0.00  & 1.11    & 0.754 & 202.91         & 0.362         & 40.94     & 43.07     & 28.11         \\
FRB20190601C & 0   & 88.52  & 28.47 & 517.0             & 0.77    & 0.12  & 0.76    & 0.175 & 223.54         & -0.235        & 39.37     & 41.80     & 28.01         \\
FRB20190601C & 1   & 88.52  & 28.47 & 502.2             & 0.77    & 0.12  & 0.76    & 0.175 & 201.91         & -0.363        & 39.36     & 41.79     & 28.03         \\
\boxed{FRB20190609A} & 1   & 345.30 & 87.94 & 600.5             & 0.69    & 0.56  & 1.02    & 0.200 & 267.97         & 0.247         & 39.81     & 42.43     & 28.59         \\
FRB20190617A & 0   & 178.60 & 83.87 & 409.9             & 0.69    & 0.76  & 1.32    & 0.065 & 290.64         & 0.146         & 38.94     & 41.41     & 28.13         \\
FRB20190617B & 0   & 56.43  & 1.16  & 459.3             & 1.14    & 0.00  & 0.96    & 0.166 & 217.37         & 0.813         & 39.47     & 41.57     & 27.20         \\
FRB20190618A & 0   & 321.25 & 25.44 & 419.3             & 0.29    & 0.38  & 0.63    & 0.068 & 149.73         & -0.290        & 38.31     & 41.08     & 28.56         \\
FRB20190625A & 0   & 227.91 & 32.88 & 400.2             & 1.76    & -0.46 & 1.08    & 0.225 & 293.38         & 0.505         & 39.80     & 41.35     & 25.89        \\
\hline
\end{tabular}
}
\end{table*}

\section{Conclusions}
\label{sec:conclusion}

Classifying FRBs as repeaters or non-repeaters has been a challenging problem. 
In this paper, we utilize unsupervised machine learning to learn the features of the FRBs in the first CHIME/FRB catalog and attempt to reveal their hidden properties. Most algorithms show classification results coincident with the groups of repeaters and non-repeaters. As a result, we draw the following conclusion:
\begin{itemize}
    \item Repeaters and non-repeaters have different distributions in many features. 
    \item Unsupervised machine learning algorithms, without the input of the observed identification of FRBs, can classify FRBs into clusters that have high correspondence with repeaters and non-repeaters. This suggests that repeaters and non-repeaters indeed belong to different categories. 
    \item In addition to spotting the two most significant categories of repeaters and non-repeaters, unsupervised learning methods also imply that there may exist 5 to 7 subspecies based on their traits.
    \item Learning from multiple parameters, unsupervised machine learning identifies some hidden repeaters from the apparently non-repeaters. Many of such candidates overlap with the candidates identified from supervised machine learning methods (Paper I, \citet{luo2022MachineLearningClassification}). These candidates can be top targets for future follow-up observations to identify more repeaters from the CHIME archives. 
\end{itemize}

\section*{Acknowledgements}

We would like to thank the UNLV transient group members Shunke Ai, Connery Chen, Emily Huerta, and Yuanhong Qu for various discussions, especially on the derivations of FRB brightness temperature. J-MZ-G acknowledges Shuqing Zhong and Zigao Dai for the equipment support and the discussions on the traits of fast radio bursts. J-WL and BZ acknowledge support from the Top Tier Doctoral Graduate Research Assistantship (TTDGRA) and Nevada Center for Astrophysics at the University of Nevada, Las Vegas. 

\section*{Data Availability}
\label{sec:data_ava}

The data we utilized in this paper from the first CHIME/FRB catalog is available at \url{https://www.chime-frb.ca/catalog}. The codes and other supplementary material are available at \url{https://github.com/JiamingZhuge/FRB_ML_unsp}.

\bibliographystyle{mnras}
\bibliography{frb-unsp-ml,frb-ml}

\appendix

\section{List of hyperparameters used}
\label{sec: hyperparameter}

\bottomcaption{List of hyperparameters we used in the dimensionality reduction, including the default values. }
\label{tab: dimension hyperparameters list}
\begin{supertabular}{ll}
		\hline
		Name & Value \\
		\hline
		\multicolumn{2}{c}{PCA} \\
		\hline
        \texttt{n\_components} & 2 \\
        \texttt{copy} & True \\
        \texttt{whiten} & False \\
        \texttt{svd\_solver} & 'auto' \\
        \texttt{tol} & 0.0 \\
        \texttt{iterated\_power} & 'auto' \\
        \texttt{n\_oversamples} & 10 \\
        \texttt{power\_iteration\_normalizer} & 'auto' \\
        \texttt{random\_state} & None \\
        \hline
		\multicolumn{2}{c}{t-SNE} \\
		\hline
		\texttt{n\_components} & 2 \\
        \texttt{perplexity} & 24 \\
        \texttt{early\_exaggeration} & 12 \\
        \texttt{learning\_rate} & 'auto' \\
        \texttt{n\_iter} & 1000 \\
        \texttt{n\_iter\_without\_progress} & 300 \\
        \texttt{min\_grad\_norm} & 1e-7 \\
        \texttt{metric} & 'euclidean' \\
        \texttt{metric\_params} & None \\
        \texttt{init} & 'random' \\
        \texttt{verbose} & 0 \\
        \texttt{random\_state} & 45 \\
        \texttt{method} & 'barnes\_hut' \\
        \texttt{angle} & 0.5 \\
        \texttt{n\_jobs} & None \\
        \texttt{square\_distances} & 'deprecated' \\
        \hline
		\multicolumn{2}{c}{UMAP} \\
		\hline
		\texttt{n\_neighbors} & 24 \\
        \texttt{n\_components} & 2 \\
        \texttt{metric} & 'euclidean' \\
        \texttt{metric\_kwds} & None \\
        \texttt{output\_metric} & 'euclidean' \\
        \texttt{output\_metric\_kwds} & None \\
        \texttt{n\_epochs} & None \\
        \texttt{learning\_rate} & 1 \\
        \texttt{init} & 'spectral' \\
        \texttt{min\_dist} & 0 \\
        \texttt{spread} & 1 \\
        \texttt{low\_memory} & True \\
        \texttt{n\_jobs} & -1 \\
        \texttt{set\_op\_mix\_ratio} & 1 \\
        \texttt{local\_connectivity} & 1 \\
        \texttt{repulsion\_strength} & 1 \\
        \texttt{negative\_sample\_rate} & 5 \\
        \texttt{transform\_queue\_size} & 4 \\
        \texttt{a} & None \\
        \texttt{b} & None \\
        \texttt{random\_state} & 54 \\
        \texttt{angular\_rp\_forest} & False \\
        \texttt{target\_n\_neighbors} & -1 \\
        \texttt{target\_metric} & 'categorical' \\
        \texttt{target\_metric\_kwds} & None \\
        \texttt{target\_weight} & 0.5 \\
        \texttt{transform\_seed} & 42 \\
        \texttt{transform\_mode} & 'embedding' \\
        \texttt{force\_approximation\_algorithm} & False \\
        \texttt{verbose} & False \\
        \texttt{tqdm\_kwds} & None \\
        \texttt{unique} & False \\
        \texttt{densmap} & False \\
        \texttt{dens\_lambda} & 2 \\
        \texttt{dens\_frac} & 0.3 \\
        \texttt{dens\_var\_shift} & 0.1 \\
        \texttt{output\_dens} & False \\
        \texttt{disconnection\_distance} & None \\
        \texttt{precomputed\_knn} & (None,None,None) \\
        \hline
\end{supertabular}

\bottomcaption{List of hyperparameters we used in the clustering, including the default values.}
\label{tab: cluster hyperparameters list}
\begin{supertabular}{ll}
		\hline
		Name & Value \\
		\hline
		\multicolumn{2}{c}{$k$-means} \\
        \hline
        \texttt{n\_clusters} & 2 \\
        \texttt{init} & 'k-means++' \\
        \texttt{n\_init} & 10 \\
        \texttt{max\_iter} & 300 \\
        \texttt{tol} & 0.0001 \\
        \texttt{verbose} & 0 \\
        \texttt{random\_state} & 4 \\
        \texttt{copy\_x} & True \\
        \texttt{algorithm} & 'lloyd' \\
        \hline
        \multicolumn{2}{c}{HDBSCAN} \\
        \hline
        \texttt{min\_cluster\_size}(t-SNE) & 32 \\
        \texttt{min\_samples}(t-SNE) & 2 \\
        \texttt{min\_cluster\_size}(UMAP) & 22 \\
        \texttt{min\_samples}(UMAP) & 8 \\
        \texttt{cluster\_selection\_epsilon} & 0 \\
        \texttt{max\_cluster\_size} & 0 \\
        \texttt{metric} & 'euclidean' \\
        \texttt{alpha} & 1 \\
        \texttt{p} & None \\
        \texttt{algorithm} & 'best' \\
        \texttt{leaf\_size} & 40 \\
        \texttt{memory} & Memory( \\
        & cachedir=None,\\
        & verbose=0)\\
        \texttt{approx\_min\_span\_tree} & True \\
        \texttt{gen\_min\_span\_tree} & False \\
        \texttt{core\_dist\_n\_jobs} & 4 \\
        \texttt{cluster\_selection\_method} & 'eom' \\
        \texttt{allow\_single\_cluster} & False \\
        \texttt{prediction\_data} & False \\
        \texttt{match\_reference\_implementation} & False \\
        \hline
\end{supertabular}

\bsp	
\label{lastpage}
\end{document}